\documentstyle[twoside,12pt]{report}
\newcommand{\bea}{\begin{eqnarray}}
\newcommand{\be}{\begin{equation}}
\newcommand{\eea}{\end{eqnarray}}
\newcommand{\ee}{\end{equation}}
\def\nn{\nonumber}

\def\a{\alpha}
\def\b{\beta}

\def\d{\delta}
\def\e{\epsilon}

\def\g{\raisebox{.4ex}{$\gamma$}}

\def\k{\kappa}
\def\l{\lambda}
\def\m{\mu}
\def\n{\nu}

\def\o{\omega}
\def\p{\pi}

\def\r{\rho}
\def\s{\sigma}

\def\F{\Phi}
\def\G{\Gamma}
\def\J{\Psi}
\def\L{\Lambda}
\def\O{\Omega}

\def\cd{{\cal D}}

\def\cg{{\cal G}}
\def\ch{{\cal H}}

\def\cl{{\cal L}}

\def\tre{\; {}^{\scriptscriptstyle{(3)}}\!}

\begin{document}
 \begin{titlepage}
\begin{flushleft}
G\"oteborg ITP 93-13\\
May 1993\\
\end{flushleft}
\vspace{5mm}
\begin{center}
{\Large Actions for Gravity, with Generalizations: A Review}\\
\vspace{5mm}
{\large Peter Peld\'{a}n}\footnote{Email address: tfepp@fy.chalmers.se}\\
\vspace{1cm}
{\sl Institute of Theoretical Physics\\
Chalmers University of Technology\\
and University of G\"oteborg\\
S-412 96 G\"oteborg, Sweden}\\
\vspace{1cm}
{\bf Abstract}\\
\end{center}
The search for a theory of quantum gravity has for a long time been almost
fruitless.
A few years ago, however, Ashtekar found a reformulation of Hamiltonian
gravity, which
thereafter has given rise to a new promising quantization project; the
canonical Dirac
quantization of Einstein gravity in terms of Ahtekar's new variables. This
project has
already given interesting results, although many important ingredients are
still
missing before we can say that the quantization has been successful.

Related to the classical Ashtekar Hamiltonian, there have been discoveries
regarding
new classical actions for gravity in (2+1)- and (3+1)-dimensions, and also
generalizations of Einstein's theory of gravity. In the first type of
generalization, one introduces infinitely many new parameters, similar to the
conventional Einstein cosmological constant, into the theory. These
generalizations are
called "neighbours of Einstein's theory" or "cosmological constants
generalizations", and the theory has the same number of degrees of freedom, per
point in spacetime, as the conventional Einstein theory. The second type is a
gauge
group generalization of Ashtekar's Hamiltonian, and this theory has the correct
number of degrees of freedom to function as a theory for a unification of
gravity
and Yang-Mills theory. In both types of generalizations, there are still
important
problems that are unresolved: {\em e.g} the reality conditions, the
metric-signature
condition, the interpretation, etc.

In this review, I will try to
clarify the relations between the new and old actions for gravity, and also
give a short
introduction to the new generalizations. The new results/treatments in this
review
are: 1. A more detailed constraint analysis of the Hamiltonian formulation of
the
Hilbert-Palatini Lagrangian in (3+1)-dimensions. 2. The canonical
transformation
relating the Ashtekar- and the ADM-Hamiltonian in (2+1)-dimensions is given. 3.
There
is a
discussion regarding the possibility of finding a higher dimensional Ashtekar
formulation.

There are also two clarifying figures (in the beginning of chapter 2 and 3,
respectively) showing the relations between different action-formulations for
Einstein
gravity in (2+1)- and (3+1)-dimensions.
\end{titlepage}

\tableofcontents
\chapter{Introduction} \pagestyle{myheadings}
\markboth{ Chapter \thechapter \ \ \ Introduction}{}
\pagenumbering{arabic}

\raisebox{-3.6mm}{\Huge T} \vspace*{-1.5mm}
he greatest challenge in theoretical physics today is to find the theory of
quantum
\hspace*{6.5mm}
gravity. That is, the union of the theory for microscopic particles, quantum
mechanics, and the theory for "cosmological" objects, the general theory of
relativity.
Despite the fact that a lot of physicists have been attacking the problem of
quantum gravity,
using a variety of different methods during a period of at least 40
years, there has been no real progress in this quest. One reason for this
failure
could be that the tools and methods used just are not adequate for this task.
Most of
the quantization attempts, so far, have treated gravity as just another
particle field
theory. There are, however, great differences between the theory of gravity and
theories describing the other fundamental forces and particles in nature. One
of the
most striking one is that in a conventional particle theory one assumes that
the
fields are propagating on a non-dynamical background spacetime, while in the
theory of
gravity it is exactly the "background" spacetime that is the dynamical field;
"the
stage is participating in the play".

Perhaps we "only" need to invent new methods specially constructed
for quantization of diffeomorphism
invariant theories, like gravity, without any need of really changing quantum
mechanics or the theory of relativity. It could, however, also be true that
already
quantum mechanics and/or the theory of relativity are wrong, so that any
attempt to
unite
these two theories will be bound to fail.

Without any experimental guidelines (as the blackbody radiation was for quantum
mechanics, and perhaps the Michelson-Morley experiment was to the special
theory of
relativity) of how to modify either our methods or our theories, we either have
to
keep on struggling with the quantization of the standard formulations, or we
could
make a "theoretical excursion", leaving the experimentally confirmed way,
and try to guess what kind of modifications our theories
need, possibly guided by theoretical beauty or other temptations.

String theory \cite{Green books} is a kind of "theoretical excursion", without
any
experimental support. It has, for the last 10 years, been regarded as the
strongest candidate for a theory of quantum gravity, or even for a theory of
everything. However, the optimism regarding string theory seems to have
decreased
slightly, recently, due to technical difficulties and absence of progress.
The basic idea
behind string theory is that the fundamental constituents of matter are
string-like,
extended, one-dimensional objects, instead of pointlike which is otherwise
believed.
In a way, string
theory can be seen as a synthesis of many earlier attempts of quantizing
gravity:
supergravity, Kaluza-Klein theories, higher spin theories, etc.

In the latest years, another way of tackling quantum gravity has received
 a lot of attention.
When Ashtekar \cite{2} in 1986 managed to reformulate the Einstein theory of
gravity in terms of new variables, it soon became clear that this new
formulation had
some appealing features that made it suitable for quantization \`{a} la Dirac.
Since then, physicists have been working on this project, and perhaps for the
first time in the history of quantum gravity there has really been some
progress
regarding solutions to the constraints of quantum gravity
 \cite{3a}, \cite{3b}. (These solutions are the physical wavefunctionals that
are
annihilated by the operator valued constraints, in the canonical formulation.)
 However, some people might say that this is not really
worth anything until we also
have an inner product and observables so that we can calculate a physical
quantity.
Of course, it could be that this critique is correct, and that these
quantization
attempts will never lead to the desired result without the introduction of
modifications somewhere. It is, however, anyhow very valuable to try the
conventional
quantization of the conventional theory first, in order to let the theory
itself
indicate in what direction we should search for modifications.

Besides these two main routes towards a theory of quantum gravity, there also
exist
attempts using path integral
quantization and numerical calculations in simplicial quantum gravity.
\\ \\
\noindent As mentioned above, the new promising quantization scheme is based on
the
new Ashtekar reformulation of Hamiltonian gravity. This Ashtekar reformulation
can
be seen as a shift of emphasis from the metric to the connection as the
fundamental field for gravity. Later, Capovilla, Dell and Jacobson (CDJ)
\cite{12}
managed to go even further and found an action written (almost) purely in terms
of
the connection. This discovery soon led to the finding of two different types
of
generalizations of both the CDJ-Lagrangian as well as the Ashtekar Hamiltonian:
the
cosmological constants in refs.\cite{19}, \cite{17} and \cite{18}, and also
the gauge group generalization in \cite{20}.

The purpose of this review is to describe
most of the known actions for classical gravity in (2+1)- and (3+1)-dimensions,
and
also to show how these different actions are related. I also want
 to briefly give the basic ideas behind the
generalizations mentioned above.

In chapter 2, I describe the actions for (3+1)-dimensional gravity, and do most
of the
calculations in great detail. Chapter 3 contains the actions for
(2+1)-dimensions, and
chapter 4 presents the generalized Ashtekar Hamiltonians.

Throughout this review I neglect surface terms that appear in partial
integrations in
the actions. That is, I assume compact spacetime or fast enough fall-off
behavior at
infinity for the fields. Note, however, that surface terms normally need a
careful
treatment \cite{Teitelboim}, \cite{Soloviev}.

My notation and conventions are partly given in each section and partly
collected in
two appendices. Definitions and notation introduced in one section are,
however, only
valid inside that section. This is specially true for the covariant derivatives
of
which, in this review, there exist at least five different types of, but only
three
different symbols for $\cd _a$, $D_a$ and $\nabla _a$.

\newpage \markboth{}{}

\begin{picture}(158,200)(8,20)

\thicklines
\put (109,20){\framebox(50,20){\shortstack{The Einstein-Hilbert\\ Lagrangian
${\cal L}(e^\alpha _I)$}}}
\put (2,20){\framebox(50,20){\shortstack{The Hilbert-Palatini\\ Lagrangian
${\cal
L}(e^\alpha _I,\omega ^{JK}_\beta)$}}}
\put (22,90){\framebox(50,20){\shortstack{Hamiltonian formulation\\ of the H-P
Lagrangian\\ ${\cal H}(\omega _a ^{IJ},\pi ^b_{KL})$}}}
\put (109,90){\framebox(50,20){\shortstack{The ADM-Hamiltonian\\ $\ch (e^a_i,\p
_b^j)$}}}
\put(60,50){\dashbox{2}(46,20){\shortstack{The pure $SO(1,3)$ spin-\\
connection
Lagrangian\\ $\cl (\o _\a ^{IJ})$}}}
\put(2,140){\framebox(50,20){\shortstack{The self dual\\ H-P Lagrangian\\ $\cl
(e^\a
_I, \o^{(+)JK}_\b)$}}}
\put (109,140){\framebox(50,20){\shortstack{The Ashtekar Hamiltonian\\ $\ch
(A_{ai},E^{bj})$\\ $\left ( A_{a i}=\o ^{(+)}_{a i}\right )$}}}
\put (2,200){\framebox(50,20){\shortstack{The Plebanski Lagrangian\\ $\cl
(\Sigma _{\a
\b}^i,\o _\d^{(+)i},\Psi^{ij},\eta)$}}}
\put (109,200){\framebox(50,20){\shortstack{The CDJ-Lagrangian\\ $\cl (\eta,
A_{\a i})$\\ $\left ( A_{\a i}=\o ^{(+)}_{\a i}\right )$}}}

\thinlines
\put (52,30){\vector(1,0){57}}
\put(134,40){\vector(0,1){50}}
\put(134,80){\vector(0,-1){40}}
\put(72,100){\vector(1,0){37}}
\put(27,40){\vector(0,1){50}}
\put(52,35){\vector(4,3){20}}
\put(65,70){\vector(0,1){20}}
\put(65,90){\vector(0,-1){20}}
\put(12,40){\vector(0,1){100}}
\put(134,110){\vector(0,1){30}}
\put(134,140){\vector(0,-1){30}}
\put(52,150){\vector(1,0){57}}
\put(27,160){\vector(0,1){40}}
\put(27,200){\vector(0,-1){40}}
\put(52,210){\vector(1,0){57}}
\put(134,160){\vector(0,1){40}}
\put(134,200){\vector(0,-1){40}}
\put(52,160){\vector(3,2){60}}

\it
\put(62,25){$\frac{\d S}{\d \o}=0 \Rightarrow \o=\o (e)$}
\put(137,62){\shortstack[l]{Legendre\\ transform}}
\put(137,83){ $\p :=\frac{\d L}{\d \dot{e}}$}
\put(137,45){$\dot{e}=\{e,H\}$}
\put(75,102){\shortstack{Solving second-\\ class constraints}}
\put(32,75){\shortstack[l]{Legendre\\ transform\\ $\p :=\frac{\d L}{\d
\dot{\o}}$}}
\put(67,75){\shortstack[l]{$\p :=\frac{\d L}{\d \dot{\o}}$\\ Legendre
transform\\
$\dot{\o}=\{\o,H\}$ }}
\put(75,40){\shortstack[l]{$\frac{\d S}{\d e}=0\Rightarrow$\\ $ e=e(\o)$}}
\put(17,130){\shortstack[l]{Keeping only\\ self dual part.}}
\put(137,120){\shortstack[l]{Complex\\ canonical\\ transform. }}
\put(100,135){$A_{ai}:=K_{ai}-i\G _{ai}$}
\put(100,112){$K_{ai}:=A_{ai}+i \G _{ai}$}
\put(137,175){\shortstack[l]{Legendre\\ transform}}
\put(137,193){$\dot{A}=\{A,H\}$}
\put(137,165){$\p :=\frac{\d L}{\d \dot{A}}$}
\put(32,180){\shortstack[l]{Letting $\Sigma$\\ become an\\ independent
\\field.}}
\put(32,163){\shortstack[l]{$\frac{\d S}{\d \Psi}=0$\\ $\Rightarrow
\Sigma=\Sigma
(e)$}}
\put(57,213){$\frac{\d S}{\d \Sigma}=0$ and $\frac{\d S}{\d \Psi}=0 \Rightarrow
$}
\put(57,205){$\Sigma=\Sigma (\o), \; \Psi=\Psi (\o)$}
\put(55,153){Legendre transform: $\p :=\frac{\d L}{\d \dot{\o}}$}
\put(90,180){\shortstack{$\frac{\d S}{\d e}=0$\\ $\Rightarrow e=e(\o^{(+)})$}}

\end{picture} \\
\begin{center}{\it Fig.1  Actions for gravity in (3+1)-dimensions}\end{center}
 \pagebreak

\chapter{Actions in  (3+1)-dimensions} \markboth{Chapter \thechapter \ \ \
Actions in
(3+1)-dimensions}{\thesection \ \ \ The Einstein-Hilbert Lagrangian}
\label{(3+1)}
\raisebox{-3.6mm}{\Huge I} \vspace*{-1.5mm}
n fig. 1, I have tried to collect most of the known classical actions for
gravity in \nolinebreak (3+1)-\hspace*{3.5mm}
dimensions, and their connecting relations. Among these actions
there is one for which the explicit form is not known, today. That is "the pure
$SO(1,3)$
spin-connection Lagrangian", and it can be found through an elimination of the
tetrad field from the Hilbert-Palatini Lagrangian or from a Legendre transform
from
the Hamiltonian formulation of the H-P Lagrangian. The reason why this
Lagrangian is
not known explicitly is due to technical difficulties in the above mentioned
calculations.
(There is, however, an "affine connection form" of this Lagrangian, that is
explicitly known; the Schr\"{o}dinger Lagrangian \cite{3c}. This Lagrangian can
be
found through an elimination of the metric from the Hilbert-Palatini
Lagrangian,
in metric and
affine connection form, and the Lagrangian equals the square-root of the
determinant of
the Ricci-tensor.)
There is also one known action that is missing in fig.1; in ref.\cite{'tHooft}
'tHooft
presented an
$SL(3)$ and diffeomorphism invariant action, and it was shown that the
equations of motion, following from this action, are the Einstein's equations.
Furthermore, if this $SL(3)$-invariance is gauge-fixed to $SU(2)$, this action
reduces
to the CDJ-action. I do not treat this action here, mainly due to the fact that
 this formulation
differs from all the actions in fig.1 in that it is constructed with the use of
an
enlarged gauge-invariance.
\section{The Einstein-Hilbert Lagrangian} \markboth{Chapter \thechapter \ \ \
Actions in
(3+1)-dimensions}{\thesection \ \ \ The Einstein-Hilbert Lagrangian} \label{EH}

In an attempt to find a Lagrangian that gives, as equations of motion, the
Einstein
equations for gravity, the obvious candidate is the simplest scalar function of
the
metric and its derivatives; the curvature scalar. And in order to get an action
which is generally coordinate invariant one must densitize the curvature scalar
with the help of the determinant of the metric. The resulting Lagrangian is
called the
Einstein-Hilbert Lagrangian for pure gravity.
In this section I will analyze this second order Einstein-Hilbert Lagrangian
for pure
gravity with a cosmological constant and show that the equations of motion
following
from the variation of the action are Einstein's equations. The Lagrangian is

\be \cl_{EH}= e\left(e^\a _I e^\b _J R_{\a \b}{}^{IJ}(\o (e)) + 2\l\right)
\label{1.1}
\ee
where $e^\a _I$ is the tetrad field and $R_{\a \b}{}^{IJ}(\o (e))$ is the
curvature of
the unique torsion-free spin-connection $\o _\a ^{IJ}$, compatible with $e^\a
_I$.
$e$ is the determinant of the inverse tetrad $e_{\a I}$. To find the change in
$S_{EH}$ under a variation of $e_{\a I}$, I need to calculate the variation of
$e,
e^\a _I, R_{\a \b}{}^{IJ}$ first:

\be e=\frac{1}{24}\e ^{\a \b \g \d}\e ^{IJKL}e_{\a I}e_{\b J}e_{\g K}e_{\d L}
\label{1.15} \ee
\be \Rightarrow \d e=\frac{1}{6}\e ^{\a \b \g \d}\e ^{IJKL}(\d e_{\a I})e_{\b
J}e_{\g K}
e_{\d L}=e e^{\a I}\d e_{\a I} \label{1.2} \ee \\

\be \d (e^\a _I e_{\a J})=(\d e^\a _I)e_{\a J}+e^\a _I \d e_{\a J}=\d
\eta_{IJ}=0 \ee

\be \Rightarrow \d e^\a _I=-e^\b _I(\d e_{\b J})e^{\a J} \label{1.3} \ee
The variation of $R_{\a \b}{}^{IJ}$ is given in Appendix B.

\be \d R_{\a \b}{}^{IJ}=\frac{1}{2}\left(e^{\g [I}\cd_{[\a}\cd_{[\b]}\d e_{\g
]}^{J]}+
e^{\g
[I}e^{\m J]}e_{K[\b}\cd_{\a]}\cd_\m \d e_{\g} ^K\right) \label{1.4} \ee
Using (\ref{1.2}), (\ref{1.3}), (\ref{1.4}) and the fact that the covariant
derivative annihilates $e_{\a I}$, the variation of $S_{EH}$ becomes:

\bea \d S_{EH}&=&\int_M d^4x\; e\left(-2 e^\a _K e^{\g} _I e^\b _J R_{\g
\b}{}^{KJ}+
(2 \l +
e^{\g} _K e^\b _J R_{\g \b}{}^{KJ})e^\a _I\right)\d e_\a ^I \nn \\ \nn \\
&+&\int_M d^4x\; \partial _\a \left(e(g^{\g [\a}e^{\b ]}_J\cd _{[\b} \d e_{\g
]}^{J]}+
g^{\g [\a} g^{\b ]\m}e_{K\b}\cd _\m \d e_{\g} ^K)\right) \label{1.5} \eea
Neglecting the surface term, the requirement that the action should be
stationary
under general variations of $e_{\a I}$, implies Einstein's equation:

\be e^\a _K e^{\g} _I e^\b _J R_{\g \b}{}^{KJ}-(\l +\frac{1}{2}e^{\g} _K e^\b
_J
R_{\g \b}{}^{KJ})e^\a _I=0 \label{1.6} \ee
This shows that the Einstein-Hilbert Lagrangian is a good Lagrangian for
gravity, in
the sense that Einstein's equation follows from its variation.

\section{The Hilbert-Palatini Lagrangian} \markboth{Chapter \thechapter  \ \ \
Actions in
(3+1)-dimensions}{\thesection \ \ \ The Hilbert-Palatini Lagrangian}

The Hilbert-Palatini Lagrangian is a first order Lagrangian for gravity which
one
gets from the Einstein-Hilbert Lagrangian simply by letting the
spin-connection, in the
argument of the Riemann-tensor, become an independent field. Normally when one
lets a
field become independent like this one needs to add a Lagrange multiplier term
to the
Lagrangian, implying the original relation, in order to get the same equations
of
motion. The reason why this is not needed here, is that the variation of the
action
with respect to the spin-connection will itself imply the correct relation.
(This is
really a
rather remarkable feature of this particular Lagrangian.)

Therefore, I consider:

\be \cl_{HP}= e\left(e^\a _I e^\b _J R_{\a \b}{}^{IJ}(\o _{\g}^{KL}) + 2
\l\right)
\label{2.1} \ee
The only difference between $\cl _{EH}$ (\ref{1.1}) and $\cl _{HP}$ (\ref{2.1})
is that
 in
(\ref{2.1}) the spin-connection $\o _{\g}^{KL}$ is an independent field, while
in
(\ref{1.1}) it is a given function of the tetrad field. When varying the action
with
respect to general variations $\d e_{\a I}$ and $\d \o_{\a}^{IJ}$, one needs to
know the
variation of $e$, $e^{\a}_I$ and $R_{\a \b}{}^{IJ}$. The two first variations
are given in
(\ref{1.2}) and (\ref{1.3}), and the variation of $R_{\a \b}{}^{IJ}$ is given
in
Appendix B.

\bea \d e&=&e e^{\a I}\d e_{\a I} \nn \\
\d e^\a _I&=&-e^\b _I(\d e_{\b J})e^{\a J} \label{2.2} \\
\d R_{\a \b}{}^{IJ}&=&\cd_{[\a}\d \o_{\b ]}^{IJ} \nn \eea
Using this together with the identity

\be e e^{[\a}_I e^{\b ]}_J=\frac{1}{2}
\e ^{\a \b \g \d}\e_{IJKL}e_{\g}^K e_\d ^L \label{2.25} \ee
gives

\bea \d S_{HP}&=&\int_M d^4x\; e \left(-2 e^\a _K e^{\g} _I e^\b _J R_{\g
\b}^{KJ}+
(2 \l +
e^{\g} _K e^\b _J R_{\g \b}^{KJ})e^\a _I\right)\d e_\a ^I \nn \\ \nn \\
&-&\int_M d^4x\;\frac{1}{2}\cd _\a\left(\e ^{\a \b \g \d} \e _{IJKL}e_{\g}^K
e_\d ^L
\right)\d
\o_{\b}^{IJ}- \partial_\a (e e^\a _I e^\b _J \d \o_\b ^{IJ}) \label{2.3} \eea
Neglecting the surface term again, the equations of motion are:

\bea e^\a _K e^{\g} _I e^\b _J R_{\g \b}{}^{KJ}-(\l +\frac{1}{2}e^{\g} _K e^\b
_J
R_{\g \b}{}^{KJ})e^\a _I&=&0 \label{2.4} \\
\cd_{[\a}e_{\g ]}^K&=&0 \label{2.5} \eea
But (\ref{2.5}) is just the zero-torsion condition, which can be solved to get
the
unique torsion-free spin-connection compatible with $e_{\a I}$. Inserting this
solution to (\ref{2.5}) into (\ref{2.4}) one gets Einstein's equation again.

This shows that the Hilbert-Palatini action also is a good action for gravity,
again
in the sense that its equations of motion are the Einstein equations. There
are,
however,
cases when the Einstein-Hilbert Lagrangian and the Hilbert-Palatini Lagrangian
gives
different theories. The first example is in a path-integral approach to quantum
gravity. In a path-integral one is supposed to vary all independent fields
freely over
all not gauge-equivalent field-configurations. And since the two Lagrangians
are only
equivalent "on-shell", and not for general field-configurations, the
path-integrals
will probably differ. The second example is gravity-matter couplings where the
spin-connection couples directly
to some matter field. This is the case for fermionic matter. In that case, the
variation of the Hilbert-Palatini action with respect to the spin-connection
will not
in general yield the torsion-free condition (\ref{2.5}). Instead the theory
will have
torsion, which, however, can be avoided by adding an extra term to the
Lagrangian. See,
for instance, \cite{4} for a recent discussion.

\section{The ADM-Hamiltonian} \markboth{Chapter \thechapter \ \ \ Actions in
(3+1)-dimensions}{\thesection \ \ \ The ADM-Hamiltonian} \label{ADM}

Here, I will derive the well-known ADM-Hamiltonian for triad gravity, starting
from the
Einstein-Hilbert Lagrangian (\ref{1.1}). The derivation will closely
follow \cite{5}.

The Einstein-Hilbert Lagrangian:

\be \cl_{EH}= e\left(R(e_{\a}^I)+2\l\right) \label{3.1} \ee
To be able to find the Hamiltonian formulation of (\ref{3.1}), I will assume
spacetime $M$ to be topologically $\Sigma \times R$, where $\Sigma$ is some
space-like
submanifold of $M$, and $R$ stands for the time direction. I will also partly
break the manifest spacetime covariance by choosing the $x^0$ coordinate to be
my
time-coordinate.

Before defining the momenta and doing the Legendre transform, it is preferable
to
slightly rewrite the Lagrangian. The reason is that (\ref{3.1}) contains
second derivatives of the tetrad, which can be partially integrated away,
leaving a
dependence solely on $e_\a ^I$ and its first derivatives.

First, I define two covariant derivatives: $\cd_\a$ and $\nabla _\a$ . $\cd
_\a$ is
covariant with respect to both general coordinate transformations in spacetime
as
well as local Lorentz transformations on the flat index, while $\nabla _\a$ is
only
covariant under general coordinate transformations. Or in other words: $\cd
_\a$
"knows how to act" on both spacetime indices as well as Lorentz indices, while
$\nabla _\a$ only "knows how to act" on spacetime indices.

\bea \cd _\a \l^{\b I}&:=&\partial _\a \l^{\b I} + \G ^\b _{\a \g} \l ^{\g I} +
\o
_{\a \; J}^I \l^{\b J} \label{3.2} \\
\nabla _\a \l^{\b I}&:=&\partial _\a \l^{\b I} + \G ^\b _{\a \g} \l ^{\g I}
\label{3.3} \eea
where $\G_{\a \g}^\b$ in (\ref{3.2}) and (\ref{3.3}) are the same connection.
Then I
require that these covariant derivatives are compatible with the tetrad and the
metric:

\bea \cd _\a e_{\b I}&=&0 \label{3.4} \\
\nabla _\a g_{\b \g}&=&\cd _\a g_{\b \g}=0 \eea
Using (\ref{3.4}) and requiring $\G _{[\a \b]}^{\g}=0$ (no torsion) it is
possible to
uniquely determine $\G _{\a \b}^{\g}$ and $\o _{\a \; J}^I$ as functions of the
tetrad. See Appendix B for details. The Riemann-tensor is then defined:

\bea R_{\a \b I}{}^J \l _J&:=&\cd _{[\a}\cd _{\b ]}\l _I \label{3.5} \\
R_{\a \b \m}{}^{\e}\l _\e&:=&\nabla _{[\a} \nabla _{\b ]} \l _\m \label{3.6} \\
\Rightarrow \cd _{[\a}\cd _{\b ]} \l _{\m I}&=&R_{\a \b \m}{}^\e \l _{\e I} +
R_{\a \b
I}{}^J \l _{\m J} \eea
Using vectors like $\l _K:=e^\a _K \l_\a$ in (\ref{3.5}) and (\ref{3.6}) it is
straightforward to show that \\ $R_{\a \b \g}{}^\d=R_{\a \b I}{}^J e_{\g}^I
e^\d _J$,
and
with the help of definition (\ref{3.6}) the
Einstein-Hilbert Lagrangian can be rewritten:

\be R(e^I _\a):=g^{\a \g}R_{\a \b \g}{}^\b =e^{\b J}\nabla_{[\a}\nabla_{\b
]}e^\a _J
\label{3.8} \ee
Then, using (\ref{3.4}), it follows that

\be \nabla _\a e^{\b J}=-\o_\a {}^{JK}e^\b _K \label{3.9} \ee
which together with (\ref{3.8}) gives

\be R(e^I _\a)=\nabla _\a (e^{[\b}_J \nabla _{\b} e^{\a ]J}) - e^\b _K
\o_{[\a}{}^{JK}
\o_{\b ]JM}e^{\a M} \label{3.10} \ee
Now, the solution to (\ref{3.4}) for
$\o _\a {}^{IJ}$ is

\be \o _\a{}^{IJ}=\frac{1}{2}e_{\a K}(\O ^{KIJ} + \O ^{JKI} - \O ^{IJK})
\label{3.11} \ee
where I have defined the anholonomy
\be \O ^{KIJ}:=e^{\a K}e^{\b I}\partial _{[\a}e_{\b ]}^J \nn \ee
Using (\ref{3.10}), (\ref{3.11}) and neglecting the surface term
the Lagrangian (\ref{3.1}) becomes:

\be \cl= e\left(\O _K{}^{LK} \O _{IL}{}^I -\frac{1}{2}\O ^{IKL}\O _{ILK}
-\frac{1}{4}\O ^{LIK}\O _{LIK} + 2\l\right). \label{3.12} \ee
Note that $\nabla _\a e=0$.
Now the Lagrangian has a form that makes it rather straightforward to do the
Legendre
transform. The Lagrangian should first be (3+1)-decomposed, and that is
achieved by
splitting the tetrad.

\be e_{0I}=N N_I+N^aV_{aI},\hspace{10mm}e_{aI}=V_{aI};\hspace{10mm}N^I
V_{aI}=0;\hspace{10mm}N^IN_I=-1 \label{3.13} \ee
This is a completely general decomposition and puts no restriction on the
tetrad. The
inverse tetrad becomes:

\be e^{0I}=-\frac{N^I}{N},\hspace{10mm}e^{aI}=V^{aI}+\frac{N^aN^I}{N};
\hspace{10mm}V^{aI}V_{bI}=\d ^a _b;\hspace{10mm}V^{aI}N_I=0 \label{3.14} \ee
And the metric gets the standard ADM-form:

\be g_{\a \b}=\left(\begin{array}{cc}
-N^2+N^aN_a& N_a\\
N_a& V_{aI}V_b ^I
\end{array} \right)\label{3.15} \ee

\be g^{\a \b}=\left(\begin{array}{cc}
-\frac{1}{N^2}& \frac{N^a}{N^2}\\
\frac{N^a}{N^2}& V^{aI}V^b _I -\frac{N^aN^b}{N^2}
\end{array} \right)\label{3.16} \ee

\bea g&:=&det(g_{\a \b})=-N^2 det(V_{aI}V_b ^I) \label{3.17} \\
e&:=&det(e_{\a I})=N \sqrt{det(V_{aI}V_b ^I)} \label{3.18} \eea
where $N_a:=V_{aI}V_b ^I N^b$. $N$ and $N^a$ are normally called the
lapse-function
and the shift-vector, respectively.

 From here, it is rather straightforward, although tedious, to perform the
Legendre
transform
using the (3+1)-decomposition (\ref{3.13}). There is, however, a simple way of
reducing
this calculation dramatically. By partly breaking the manifest Lorentz
invariance
with the gauge-choice $N_I=(1,0,0,0)$, the Legendre transform simplifies a lot.
And
since the only difference at the Hamiltonian level should be that the Lorentz
invariance is reduced to $SO(3)$ invariance, I use this gauge-choice here. With
this
choice for $N^I$ there is no inconsistency in notation with $V^{ai}$ called
$e^{ai}$,
so I change notation to: $e^{ai}=V^{ai},\hspace{5mm}e_{ai}=V_{ai}$.

Now, I use the (3+1)-decomposition of $e^{\a I}$ for decomposing $\O ^{IJK}$
also.

\bea \O^{0jk}&=&\frac{1}{N}e^{bj}\partial _{[0}e_{b]}^k
-\frac{N^a}{N}e^{bj}\partial
_{[a}e_{b]}^k \nn \\
\O ^{0j0}&=&\frac{1}{N}e^{bj}\partial _b N \label{3.19} \\
\O ^{ij0}&=&0 \nn \\
\O ^{ijk}&=&e^{ai}e^{bj}\partial _{[a}e_{b]}^k \nn \eea
The Lagrangian (\ref{3.12}) should also be (3+1)-decomposed:

\bea \cl &=& N\;{}^{(3)}e\left (-\frac{1}{2}\O ^{0(kl)}\O _{0kl}+ \O ^0{}_k{}^k
\O _{0j}{}^j+2\O ^i{}_0{}^0 \O _{ij}{}^j \right. \nn \\
&&\left. -\frac{1}{4}\O ^{ijk}\O _{ijk} -\frac{1}{2}\O ^{ijk}\O _{ikj}
+\O _k{}^{ik}\O _{ji}{}^j+2\l\right ) \label{3.20} \eea
where ${}^{(3)}e:=det(e_{ai})$.
By using (\ref{3.10}), but now for the three-dimensional covariant derivative
$\nabla _a$
on $\Sigma $, defined to annihilate $e_{ai}e_b^i$, one can easily show that:

\bea N\tre e\tre R&=&N\tre e\left (\O _k{}^{ik}\O _{ji}{}^j-\frac{1}{4}
\O ^{ijk}\O _{ijk} -\frac{1}{2}\O ^{ijk}\O _{ikj}\right) \nn \\
&&+\nabla _{[a}(N\tre ee^{bk}\nabla_{b]}e^a_k)-2\tre ee^b_i\partial_
{[b}e^i_{c]}g^{ca}\partial _a N \label{3.21} \eea
Using this in (\ref{3.20}) and again neglecting surface terms gives

\be \cl = N\tre e\left (-\frac{1}{2}\O ^{0(kl)}\O _{0kl}+ \O ^0{}_k{}^k \O
_{0j}{}^j +
\tre R + 2\l\right) \label{3.22} \ee
where I have used (\ref{3.19}) to eliminate the last term in (\ref{3.21}). Now
it is
time to define the momenta

\be \p ^{ai}:=\frac{\d L}{\d \dot{e}_{ai}}=\tre e(\O ^{0(ki)}e^a_k - 2 \O
^0{}_k{}^k e^{ai}) \label{3.23} \ee
which means that there exists a primary constraint:

\be L^i:=\e ^{ijk}\p ^a_j e_{ak}\approx 0 \label{3.24} \ee
It will be shown later that this constraint is the generator of $SO(3)$
rotations.
Inversion of (\ref{3.23}) yields:

\be \O ^{0(ki)}=\frac{1}{{}^{\scriptscriptstyle{(3)}}e}(\p
^{ai}e_a^k-\frac{1}{2}
(\p ^{bj}e_{bj})\d ^{ik})
\label{3.25} \ee
Altogether, the total Hamiltonian becomes:

\bea \ch&=&\p ^{ai}\dot{e}_{ai}-\cl + \L _i L^i =\frac{N}{2\tre e}\left(\p
^{ij}\p_{ij}
-\frac{1}{2}(\p ^i{}_i)^2 -2\tre e^2\tre R-4\tre e^2\; \l\right) \nn \\
&& -N^ce_{ci}\tre \cd _a\p ^{ai} + \L _iL^i \label{3.26} \eea
where $\p ^{ij}:=\p ^{ai}e^j_a$, $\tre \cd _a$ is the covariant derivative
defined to
annihilate $e_{ai}$, and $\L _i$ is an arbitrary Lagrange multiplier.
The fundamental Poisson bracket is

\be \{ e_{ai}(x),\p ^{bj}(y)\} =\d ^b _a \d ^j _i \d ^3(x-y) \label{3.265} \ee
The geometrical interpretation of these phase space variables is that $e_{ai}$
is
the triad field on the spatial hypersurface, and $\p ^{bj}$ is closely related
to
the extrinsic curvature of the hypersurface.
Since $N$ and $N^a $ have vanishing momenta the variation of them implies the
secondary constraints:

\bea \ch &:=&\frac{1}{2\tre e}\left(\p ^{ij}\p_{ij}
-\frac{1}{2}(\p ^i{}_i)^2 -2\tre e^2\tre R-4\tre e^2\; \l\right)\approx 0
\label{3.27}
\\
\ch _a &:=&-e_{ai}\tre \cd _b \p ^{bi} \approx 0 \label{3.28} \eea
normally called the Hamiltonian- and the vector-constraint.

Then, in order to check if the Hamiltonian (\ref{3.26}) is the complete
Hamiltonian
for this theory one needs to calculate the time evolution of the constraints.
The
requirement is that the time evolution of the constraint must be weakly zero.
For
a total Hamiltonian which is just a linear combination of constraints, this
corresponds to requiring the constraints to be first-class. To calculate the
constraint
algebra, I will first show that the transformations generated by $L^i$ are
$SO(3)$
rotations, and the ones generated by $\ch _a$ are spatial diffeomorphisms
modulo
$SO(3)$ rotations. The transformations generated by $L^i$:

\bea \d ^{L^i}e_{ai}&=&\{e_{ai},L^i[\L _i]\} = -\e_{ijk}\L ^j e^k_a
\label{3.285} \\
 \d ^{L^i}\p ^{a}_i&=&\{\p ^{a}_i,L^j[\L _j]\} = -\e_{ijk}\L ^j \p ^{ka}
\label{3.286}
\eea
where $L^i[\L _i]:=\int _\Sigma d^3x\; L^i(x)\L _i(x)$. This shows that $L^i$
is
the generator of $SO(3)$ rotations. Then, I will define the new
constraint $\tilde{\ch }_a$ as a linear combination of $L^i$ and $\ch _a$, and
show
that the transformations generated by this new constraint are spatial
diffeomorphisms:

\be  \tilde{\ch }_a:=\ch _a+M_{ai}(e_{bj})L^i \label{3.31} \ee
where

\be M_{ci}(e_{bj}):=\frac{1}{2}\e_{ijk}\tre \o_c^{jk}=
\frac{1}{2}\e _{ijk}e^{aj}e^{bk}(e_{bl}\partial _{[a}e_{c]}^l +
e_{cl}\partial _ae_b^l). \nn \ee
The transformations of the fundamental fields are:

\bea \d ^{\tilde{\ch}_a}e_{ai}\! &=&\! \{e_{ai},\tilde{\ch }_b[N^b]\}\! =\!
N^b\partial _be_{ai}\! +\!
e_{bi}\partial _aN^b\!=\! \pounds _{N^b}e_{ai} \label{3.29} \\
\d ^{\tilde{\ch}_a}\p ^{ai}\! &=&\! \{\p ^{ai},\tilde{\ch }_b[N^b]\}\! =\! N^b
\partial _b \p ^{ai}\!
-\! \p ^{bi}\partial _b N^a \! +\! \p ^{ai}\partial _bN^b\! =\! \pounds
_{N^b}\p ^{ai}
\label{3.30}
\eea
where $\pounds _{N^b}$ denotes the Lie-derivative in the direction $N^b$. The
last
term in (\ref{3.30}) is needed since $\p ^{ai}$ is a tensor density of
weight plus one. With the information that $L^i$ generates $SO(3)$ rotations
and
$\tilde{\ch}_a$ generates spatial diffeomorphisms, it is really simple to
calculate
all Poisson brackets containing these two constraints. All one needs to know is
how
the constraints transform under these transformations. Under $SO(3)$ rotations,
$L^i$
is a vector and $\ch $ is a scalar. Under spatial diffeomorphisms, $L^i$ and
$\ch $
are scalar densities, and $\tilde{\ch }_a$ is a covariant vector density. (Note
that
$\tilde{\ch }_a$ is {\it{not}} an $SO(3)$ scalar since $K_{ai}$ does not
transform
covariantly under $SO(3)$ transformations. Note also that due to the
anti-symmetry in
the derivatives in $K_{ai}$, $K_{ai}$ and $\tilde{\ch}_a$ do, however,
transform covariantly under spatial diffeomorphisms.) This gives the algebra:

\bea \{L^i[\L _i],L^j[\g _j]\}&=&L^k[\e _{kij}\g ^i\L ^j] \nn \\
\{L^i[\L _i],\tilde{\ch}_a[N^a]\}&=&L^i[-\pounds _{N^a}\L _i] \nn \\
\{L^i[\L _i],\ch [N]\}&=&0 \label{3.34} \\
\{\tilde{\ch}_a[N^a],\tilde{\ch}_b[M^b]\}&=&\tilde{\ch}_a[\pounds _{N^b}M^a]
\nn \\
\{\ch [N],\tilde{\ch}_a[N^a]\}&=&\ch [-\pounds _{N^a}N] \nn \eea
The only Poisson bracket left to calculate is the one including two Hamiltonian
constraints: $\{\ch [N],\ch [M]\}$. This calculation is a bit trickier, but
since the
result must be anti-symmetric in $N$ and $M$, one can neglect all terms not
containing
derivatives on these fields. Then, one needs the variation of $\tre R$ with
respect to
$e_{ai}$:

\be \frac{\d \tre R(e_{bj}(y))}{\d e_{ai}(x)}=\cd _b\left (g^{a[b}(e^{d]i}\cd
_d\d
^3(x-y)+g^{d]e}e_{d}^i\cd _e\d ^3(x-y))\right ) \label{3.35} \ee
A straightforward calculation gives

\bea \{\ch [N],\ch [M]\}&=&\int_{\Sigma}d^3\! z\int_{\Sigma}d^3\! x\left
(-N(x)\tre
e(x)
\frac{\d \tre R(x)}{\d e_{ai}(z)}\frac{M(z)}{\tre e(z)}\right. \nn \\&&\times
\left. (\p
^{ij}(z)e_{aj}(z)-\frac{1}{2}\p ^k{}_k(z)e_a^i(z))\right ) - (N\leftrightarrow
M)\nn \\
&=&\cdots =\tilde{\ch}_a[g^{ab}(N\partial _bM-M\partial _bN)] \label{3.36} \eea
where I have neglected terms proportional to $L^i$. This shows that the set of
constraints $\ch, \ch _a$ and $L^i$ really forms a first class set, meaning
that the
total Hamiltonian (\ref{3.26}) is complete and consistent.

The Hamiltonian (\ref{3.26}) is the well-known ADM-Hamiltonian for gravity,
\cite{6}.
 In attempts at canonical quantization of this Hamiltonian, the complicated
non-polynomial form of the Hamiltonian constraint has always been a huge
obstacle.
 No one has yet found an explicit
solution to the quantum version of $\ch =0$.

\section{Hamiltonian formulation of the H-P Lagrangian} \markboth{Chapter
\thechapter
\ \ \ Actions in
(3+1)-dimensions}{\thesection \ \  \ Hamiltonian formulation of the H-P
Lagrangian}
\label{HHP}

In this section, I will perform the Legendre transform from the
Hilbert-Palatini
Lagrangian. The basic canonical coordinate will here be the $SO(1,3)$
spin-connection,
and the metric will only be a secondary object expressible in terms of the
canonical
fields. The straightforward Hamiltonian analysis of this Lagrangian reveals
that there
are ten first class constraints, generating the local symmetries of the
Lagrangian
(diffeomorphisms and Lorentz transformations), and also twelve second class
constraints. The second class constraints are needed since the $SO(1,3)$
connection
has too many independent components in comparison with the tetrad field.

After deriving the complete and consistent Hamiltonian for this system, I will
go on
and solve the second class constraints, and show that the resulting Hamiltonian
is the
$SO(1,3)$ ADM-Hamiltonian. Due to the appearance of tertiary second-class
constraints,
in this analysis, the calculations in this section will be rather lengthy and
tedious.

This Legendre transform has already been done in \cite{7}
and also in \cite{8}.
The differences in my approach are that I give the
constraint algebra in more detail, and also that I solve the second class
constraints
without breaking the manifest $SO(1,3)$ invariance.

The first order Lagrangian density is

\be \cl=e(e^\a _Ie^\b _JR_{\a \b}{}^{IJ}(\o_{\g} ^{KL}) + 2\l) \label{4.1} \ee
where $e^\a _I$ is the tetrad field, $R_{\a \b}{}^{IJ}(\o_{\g} ^{KL})$ is the
curvature
of the $SO(1,3)$ connection $\o _{\g} ^{KL}$, here treated as an independent
field.
The (3+1)-decomposition gives

\be \cl=e(2e^0 _I e^a _J R_{0a}{}^{IJ} + e^a _I e^b _J R_{ab}{}^{IJ} + 2\l)
\label{4.2} \ee
I define the momenta conjugate to $\o _a ^{IJ}$:

\be \p ^a _{IJ}:=\frac{\d L}{\d \dot{\o}_a ^{IJ}}=ee^0 _{[I}e^a _{J]}
\label{4.3} \ee
Since $\p ^a _{IJ}$ has 18 components, while the right hand side of (\ref{4.3})
has only
 12 independent components, there are six primary constraints:

\be \F ^{ab}:=\frac{1}{2}\p ^a _{IJ} \p ^b _{KL}\e^{IJKL}\approx 0 \label{4.4}
\ee
The Lorentz-signature condition on the metric also imposes the non-holonomic
constraint: \\ $det(\p ^a _{IJ}\p ^{bIJ})<0$. For a discussion of the
equivalence of
(\ref{4.3}) and (\ref{4.4}) together with the non-holonomic constraint, see
\cite{8}.

Now, I want to express the Lagrangian in terms of the coordinate $\o _a ^{IJ}$,
the
momenta $\p ^a _{IJ}$, and possible Lagrange multiplier fields, and to do that
I make
use of the general decomposition of the tetrad given in (\ref{3.14}),

\be e^{0I}=-\frac{N^I}{N},\hspace{10mm}e^{aI}=V^{aI}+\frac{N^aN^I}{N};
\hspace{5mm}N^I V^a _I =0;\hspace{5mm} N^I N_I=-1\label{4.5} \ee
Using this decomposition it is possible to show that

\be \eta ^{IJ}=-N^IN^J + V^{aI}V_a ^J:=- N^IN^J+
\tilde{\eta} ^{IJ} \label{4.6} \ee
where $V_a ^I:=V_{ab}V^{bI}$ and $V_{ab}V^{bI}V^c _I=\d ^c _a$. This formula
can then
be used to project out components parallel and normal to $N^I$. With the use of
this,
and the fact that

\be \F ^{ab}=0 \Rightarrow \p^a _{IJ}\tilde{\eta}^{IK}\tilde{\eta}^{JL}=0 \ee
it is straightforward to rewrite the terms in the Lagrangian as:

\bea e\; e^a _I e^b _JR_{ab}{}^{IJ}&=&-\frac{N^2}{e}Tr(\p ^a \p ^b R_{ab})+
N^aTr(\p ^b R_{ab}) - \l
_{ab}\F ^{ab} \label{4.7} \\
e&=&det(e_{\a I})=N\sqrt{det(V_{ab})}=\frac{N^2}{e}\sqrt{det(Tr(\p ^a \p ^b))}
\label{4.8} \eea
where I have included all terms containing $\p^a
_{IJ}\tilde{\eta}^{IK}\tilde{\eta}^{JL}$
into the Lagrange multiplier term $\l _{ab}\F ^{ab}$.
Redefining the lapse function $\tilde{N}:=\frac{N^2}{e}$, the total Lagrangian
becomes

\be \cl=-Tr(\p ^{a}\dot{\o}_{a})-Tr(\o _{0}\cg) -\tilde{N}\ch -N^a\ch _a -\l
_{ab}\F ^{ab} \label{4.9} \ee
where

\bea \cg ^{IJ}&:=&\cd _a \p ^{aIJ}=\partial _a \p ^{aIJ} + [\o _a,\p
^a]^{IJ}\approx0 \nn \\
\ch&:=&Tr(\p ^a \p ^b R_{ab})-2\l \sqrt{det(Tr(\p ^a \p ^b))}\approx0 \nn \\
\ch _a&:=&-Tr(\p ^b R_{ab})\approx0 \label{4.10} \\
\F ^{ab}&:=&\frac{1}{2}\p ^{aIJ}\p ^{bKL}\e_{IJKL}\approx0 \nn \eea
and the fundamental Poisson bracket is

\be \{\o _{aIJ}(x),\p^{bKL}(y)\}=\frac{1}{2}\d ^b _a \d ^{[K}_I \d^{L]}_J\d
^3(x-y) \ee
The $SO(1,3)$ trace is defined as:
\bea Tr(ABC)&:=&A_I{}^JB_J{}^KC_K{}^I \nn \\ Tr(AB)&:=&A_I{}^J
B_J{}^I \nn \eea
The preliminary Hamiltonian can now be read off in (\ref{4.9}), using: $\ch
_{tot}=-Tr(\p
^a \dot{\o}_a) -\cl$. The Lagrange multiplier fields are $\tilde{N}, N^a, \o
_{0IJ}$
and $\l _{ab}$, and their variation imposes the constraints in (\ref{4.10}).
For this
Hamiltonian to be a complete and consistent one, one needs to check if the time
evolution of all the constraints vanishes weakly. That is, a
field-configuration that
initially satisfies all the constraints must stay on the constraint surface
under the
time evolution. In our case, the total Hamiltonian is just a linear combination
of
constraints, which means that the above requirement corresponds to demanding
that the
constraint algebra should close. If it is not closed, one has two options;
either
one fixes
some of the Lagrange multipliers so that the time evolution is consistent, or
one
introduces secondary constraint.

To simplify the constraint analysis, I prefer to change to vector notation in
$SO(1,3)$ indices.

\be \p ^{aIJ}:=\p^{ai}T_i^{IJ};\hspace{10mm}\o _a ^{IJ}:=\o_a ^i T_i ^{IJ} \nn
\ee
where $T_i^{IJ}$ are the generators of the $so(1,3)$ Lie-algebra. {\em N.B},
the indices; $i,\;j\;k$ take values $1,\; 2 \dots ,6$ here since $so(1,3)$ is a
six-dimensional Lie-algebra, while in other sections $i,\;j,\;k$ denote $SO(3)$
indices. The following
definitions and identities are also useful:

\bea q_{ij}&:=&-Tr(T_i T_j)\hspace{10mm}q^{ik}q_{kj}=\d ^i _j \nn \\
q^*_{ij}&:=&-Tr(T^*_iT_j):=-\frac{1}{2}\e^{IJKL}T_{iKL}T_{jJI} \nn \\
\p^a _i&:=&\p ^{aj}q_{ij};\hspace{10mm} \p ^{*a}_i:=\p ^{aj}q^*_{ij}
\label{4.11} \\
q^{*ij}&:=&q^{ik}q^{jl}q^*_{kl}\Rightarrow\hspace{10mm}q^{*ij}q^*_{jk}=-\d ^i
_k \nn \\
{[}T_i,T_j{]}&=&f_{ij}{}^kT_k;\hspace{10mm}f_{ijk}:=f_{ij}{}^lq_{kl}=-2Tr(T_iT_jT_k) \nn
 \\
f^*_{ijk}&:=&f_{ij}{}^lq^*_{kl} \nn \\
f_{ijk}f_{lm}{}^k&=&\frac{1}{2}(q_{i[l}q_{m]j}-q^*_{i[l}q^*_{m]j}) \nn \eea
$f^*_{ijk}$ has all the index-symmetries (anti-symmetries) that $f_{ijk}$ has.
The
$*$-operation really corresponds to the duality-operation in $so(1,3)$ indices.
Using this
notation, the Hamiltonian becomes

\bea \ch _{tot}&=&\tilde{N}\ch +N^a\ch _a +\l_{ab}\F ^{ab} -\o _{0i}\cg ^i
\label{4.12}\\
\cg ^i&=&\cd _a \p ^{ai}=\partial _a \p ^{ai}+f^i{}_{jk}\o _a ^j \p
^{ak}\approx0 \nn
\\
\ch&=&-\frac{1}{2}f_{ijk}\p ^{ai}\p ^{bj}R_{ab}{}^k-2\l \sqrt{-det(\p ^{ai}\p
^b
_i)}\approx0\nn \\
\ch _a&=&\p ^{bi}R_{abi}\approx0 \nn \\
\F^{ab}&=&\p ^{ai}\p ^{*b}_i\approx0\nn \eea
where $"\approx "$ denotes weak equality, which means equality on the
constraint-surface, or equality modulo the constraints.

The fundamental Poisson bracket is

\be \{\o _{ai}(x),\p ^{bj}(y)\}=\d ^b _a \d ^j _i \d^3(x-y) \ee
To calculate the constraint algebra I will again use the shortcut described in
section \ref{ADM}. First, the generator of spatial diffeomorphisms and the
generator
of Lorentz transformations are identified. Then, using their fundamental
transformations all Poisson brackets containing these generators are easily
calculated.

The transformations generated by $\cg ^i$ are:

\bea \d ^{\cg ^i}\p ^{ai}&:=&\{\p ^{ai},\cg ^j[\L_j]\}=f^i{}_{jk}\L^j\p ^{ak}
\label{4.14} \\
\d ^{\cg ^i}\o _a ^i&=&\{\o _a ^i,\cg ^j[\L_j]\}=-\cd _a\L^i \label{4.15} \eea
which shows that $\cg ^i$ is the generator of $SO(1,3)$ transformations. Then I
define

\be \tilde{\ch}_a:=\ch _a-\o _{ai}\cg ^i \label{4.16} \ee
and calculate the transformations it generates:

\bea \d ^{\tilde{\ch}_a}\p ^{ai}&:=&\{\p
^{ai},\tilde{\ch}_b[N^b]\}\!=\!N^b\partial _b
 \p
^{ai} \!-\! \p ^{bi}\partial _b N^a \!+\! \p ^{ai} \partial _b N^b\!=\!
\pounds _{N^b}\p ^{ai}
\label{4.17} \\
\d ^{\tilde{\ch}_a}\o _a ^i&=&\{\o _a ^i,\tilde{\ch}_b[N^b]\}\!=\!N^b\partial
_b
\o _a ^i \!+\!
\o _{b}^i\partial _a N^b\!=\!\pounds _{N^b}\o _a ^i \label{4.18} \eea
which shows that $\tilde{\ch}_a$ is the generator of spatial diffeomorphisms.
($\pounds
_{N^b}$ is the Lie-derivative along the vector field $N^b$.) This makes it easy
to
calculate all Poisson brackets containing $\cg ^i$ or $\tilde{\ch}_a$. See
section
\ref{ADM} for details.

\bea \{\cg ^i[\L _i],\cg ^j[\g _j]\}&=&\cg _i[\L ^k\g ^j f_{kj}{}^i]
\label{4.19} \\
\{\ch [N],\cg ^j[\g _j]\}&=&0 \label{4.20} \\
\{\F ^{ab}[\l _{ab}],\cg ^j[\g _j]\}&=&0 \label{4.21} \\
\{\cg ^i[\L _i],\tilde{\ch}_a[N^a]\}&=&\cg _i[-\pounds _{N^a}\L ^i]
\label{4.22} \\
\{\tilde{\ch}_b[M^b],\tilde{\ch}_a[N^a]\}&=&\tilde{\ch}_b[-\pounds _{N^a}M^b]
\label{4.23} \\
\{\ch [N],\tilde{\ch}_a[N^a]\}&=&\ch [-\pounds _{N^a}N] \label{4.24} \\
\{\F ^{ab}[\l _{ab}],\tilde{\ch}_a[N^a]\}&=&\F ^{ab}[-\pounds _{N^a}\l _{ab}]
\label{4.25} \eea
This leaves only three Poisson brackets left to calculate: \small $\{\ch
[N],\ch [M]\},
 \{\ch
[N],\F^{ab}[\l _{ab}]\}$ \normalsize and \small $\{\F ^{ab}[\l _{ab}],\F
^{cd}[\g _
{cd}]\}$.
\normalsize They are
rather straightforward to calculate, but it simplifies to notice that $\{\ch
[N],
\ch [M]\}$ is anti-symmetric in $N$ and $M$ meaning that it is only terms
containing
derivatives on these fields that survive. One needs also the structure-constant
identity in (\ref{4.11}).

\bea \{\ch [N],\ch [M]\}&=&\ch _b[\frac{1}{2}\p ^{bj}\p ^c _j(M\partial
_cN-N\partial
_cM)]\nn \\
&&+\F ^{bc}[\p ^{*am}R_{bam}(N\partial _cM-M\partial _cN)] \label{4.26}
\\
\{\F ^{ab}[\l _{ab}],\F ^{cd}[\g _{cd}]\}&=&0 \label{4.27} \\
\{\F^{ab}[\l _{ab}],\ch [N]\}&=&\int _\Sigma d^3\! x\; N\l _{ab}\p
^{*mc}f_{min}\p
^{bi}\cd _c\p ^{na}\nn \\
&&:=\J ^{ab}[N\l _{ab}] \label{4.28} \eea
The constraint algebra fails to close due to (\ref{4.28}). Notice that, if we
should simply remove the constraint $\F ^{ab}\approx0$ from the theory, the
constraint
algebra would still fail to close, this time due to (\ref{4.26}). There is,
however, an
alternative strategy that really gives a closed constraint algebra: forget
about $\F
^{ab}$, and define instead two new constraints $\ch ^*:=\frac{1}{2}f^*_{ijk}\p
^{ai}\p ^{bj}R_{ab}{}^k\approx 0$ and $\ch ^* _a:=\p ^{*bi}R_{abi}\approx 0$.
This
will make the constraints $\cg ^i,\; \ch ,\; \ch ^*,\; \ch _a$ and $\ch ^* _a$
form a first class set. The theory will of course not be ordinary Einstein
gravity,
more like twice the theory of Einstein gravity. (The number of degrees of
freedom is
four per spacetime point, and if one splits the Euclidean theory into self dual
and
anti-self dual parts, the action will really just be two copies of the pure
gravity
action. The Lorentzian case is more complicated due to the reality conditions.)

Now, returning to the theory that really follows from the H-P Lagrangian we
notice
that since the constraint algebra fails to close, the time evolution of the
constraints $\ch$ and $\F ^{ab}$ will not automatically be consistent. That is,
the
time evolution will bring the theory out of the solution-space to these
constraints.
This must be taken care of in order to get a fully consistent theory. And as
mentioned earlier, there are two different strategies available; either solve
for some
Lagrange multiplier, or introduce secondary constraints. The first method is
often the
preferred one since that solution maximizes the number of degrees of freedom.
I try this method first. The equations that should be solved are

\bea \dot{\F}^{ab}[K_{ab}]&=&\{\F ^{ab}[K_{ab}],H_{tot}\}\approx
\J^{ab}[\tilde{N}K_{ab}]\approx 0
\label{4.291} \\
\dot{\ch}[M]&=&\{\ch [M],H_{tot}\}\approx \J^{ab}[-M\l_{ab}]\approx 0
\label{4.292}
\eea
Here, $K_{ab}$ and $M$ are arbitrary test functions, and $\tilde{N}$ and $\l
_{ab}$
are the
Lagrange multiplier fields that sit in the total Hamiltonian (\ref{4.12}).
These
equations must be solved for the Lagrange multipliers so that they are
satisfied
for all test-functions $M$ and $K_{ab}$.
Assuming $\J ^{ab}$ to be a non-degenerate matrix, the minimal
solution is

\bea \tilde{N}&=&0 \label{4.295} \\
\l _{ab}&=&\tilde{\l }_{ab}-\frac{1}{3}\J _{ab}\J ^{cd}\tilde{\l}_{cd}
\label{4.296}
\eea
where $\J _{ab}$ is the inverse to $\J ^{ab}$, and $\tilde{\l}_{cd}$ is an
arbitrary
Lagrange multiplier. In a generic field theory, this would probably be a good
choice
to get a consistent Hamiltonian formulation. In our case, however, we do know
that the
Lagrange multiplier field $\tilde{N}$ has an important physical interpretation,
it is
the lapse-function, and putting that to zero would mean that the spacetime
metric
always would be degenerate. Before leaving this "unphysical" solution let us
calculate the physical degrees of freedom: half the phase space coordinates
(18) minus
the number of first class constraints (14) minus half the number of second
class
constraints (1), gives three degrees of freedom per point. Thus, it looks like
the
"degenerate metric theory" has more freedom than the conventional
non-degenerate one.
(If this "degenerate theory" is liberated from second-class constraints, and if
also
all first-class constraints except $\ch _a$ and the rotational part of $\cg ^i$
are
solved and properly gauge-fixed, then we would possibly have an $SO(3)$ and
diff($\Sigma$)-invariant theory on $SO(3)$ Yang-Mills phase space: the
Husain-Kucha\v{r}-theory \cite{Husain}. Although, to really be sure that this
"degenerate theory" is the Husain-Kucha\v{r} theory, one must show that this
reduction
can be done so that the remaining phase space coordinate is an $SO(3)$
connection\footnotemark.)\footnotetext{I thank Ingemar Bengtsson for suggesting
this.}

Excluding this solution, we are forced to include secondary constraints. The
new
constraint, which should be added to the total Hamiltonian, is

\be \J ^{ab}:=\frac{1}{2}\p ^{*mc}f_{min}\p ^{i(b}\cd _c\p ^{a)n}\approx0
\label{4.30}
\ee
meaning that the total Hamiltonian now becomes

\be \ch'_{tot}=\ch _{tot}+\g _{ab}\J ^{ab} \label{4.31} \ee
With this new Hamiltonian, one needs to check again if the time evolution of
all
constraints vanishes weakly, and the new ingredients are all Poisson brackets
containing the constraint $\J ^{ab}$. The Poisson brackets containing
$\tilde{\ch}_a$
and $\cg ^i$ are again easy to calculate. Note that, due to the fact that $\cd
_a$ is
only covariant with respect to $SO(1,3)$ transformations, $\J ^{ab}$ is not
manifestly
covariant under spatial diffeomorphisms. It is, however, covariant, which can
be seen by
adding a fiducial affine connection to $\cd _a$ and note that it drops out from
$\J
^{ab}$. (This is of course also true for $\cg ^i$.) Thus, since $\J ^{ab}$
transforms
covariantly under both $SO(1,3)$ transformations as well as spatial
diffeomorphisms,
the
Poisson brackets are

\bea \{\J ^{ab}[\g _{ab}],\cg ^i[\L _i]\}&=&0 \label{4.32} \\
\{\J ^{ab}[\g _{ab}],\tilde{\ch}_c[N^c]\}&=&\J ^{cd}[-\pounds _{N^a}\g _{cd}]
\label{4.33} \eea
The Poisson bracket between $\J ^{ab}$ and $\F ^{ab}$ is also easily
calculated:

\be \{\F ^{ab}[\l _{ab}],\J ^{cd}[M_{cd}]\}\approx \int _\Sigma d^3\! x\;
(-\frac{1}{2}
\l _{ab}\p
^{bc}M_{cd}\p ^{da}+\frac{1}{2}\l _{ab}\p ^{ab}M_{cd}\p ^{cd}) \label{4.34} \ee
where $\p ^{ab}:=\p ^{ai}\p ^b_i$. This shows that $\F ^{ab}$ and $\J ^{ab}$
are
second-class constraints. (The rank of the "constraint-matrix" in (\ref{4.34})
is, for
generic field-configurations, maximal, i.e rank twelve.) Now, two more
Poisson brackets should be calculated, $\{\J^{ab},\J^{cd}\}$ and
$\{\ch,\J^{ab}\}$. But these calculations are really horrible, and since I do
not need
the exact result in order to show that the Hamiltonian now can be put in a
consistent form, I will not write out these Poisson brackets here. Instead, I
start
with the time evolution of $\F^{ab}$:

\be \dot{\F}^{ab}[\r _{ab}]=\{\F^{ab}[\r _{ab}],H'_{tot}\}\approx\int _\Sigma
d^3\! x\;
 (-\frac{1}{2}\r _{ab}\p
^{bc}\g _{cd}\p ^{da}+\frac{1}{2}\r _{ab}\p ^{ab}\g _{cd}\p ^{cd})\approx0
\label{4.35} \ee
And since this equation is required to be satisfied for all test functions $\r
_{ab}$, the only solution is

\be \g _{ab}=0 \label{4.355} \ee
meaning that the constraint $\J^{ab}$ drops out from the total Hamiltonian.
With this
solution for $\g _{ab}$, the time evolution for $\ch$ is automatically okey and
it is
only $\J^{ab}$'s time evolution left to check. First, I define

\be \{\J^{ab}[K_{ab}],\ch [\tilde{N}]\}:=\Sigma^{ab}[K_{ab}] \label{4.37} \ee
Whatever complicated result the above Poisson bracket gives, I partially
integrate it
so that the test function $K_{ab}$ is free from derivatives, and then I call
the
result $\Sigma ^{ab}$. With this definition, the time evolution of $\J ^{ab}$
becomes

\bea \dot{\J}^{ab}[K_{ab}]=\{\J^{ab}[K_{ab}],H_{tot}\}&\approx&\int _\Sigma
d^3\! x\;
 (\frac{1}{2}K_{ab}\p
^{bc}\l _{cd}\p ^{da}-\frac{1}{2}K_{ab}\p ^{ab}\l _{cd}\p ^{cd})\nn \\
&&+ \Sigma ^{ab}[K_{ab}]\approx0
\label{4.38} \eea
This equation can always be solved for $\l _{ab}$ so that the equation is
satisfied for
all test-functions $K_{ab}$:

\be \l_{ab}=2\p _{ae}\Sigma ^{ec}\p _{cb} -\p _{ab}\p_{ce}\Sigma
^{ce} \label{4.39} \ee
Here, $\p _{ab}$ is the inverse to $\p ^{ab}:=\p^{ai}\p^b_i$.
So, regardless of the complicated form of $\Sigma^{ab}$ one can always solve
for $\l
_{ab}$. Note that it can happen that $\Sigma ^{ab}$ contains derivatives of the
lapse-function $\tilde{N}$, but in the action these can always be partially
integrated
away to get a multiplicative Lagrange multiplier.

To summarize the Hamiltonian analysis; the total and consistent Hamiltonian is

\be \ch_{tot}=\tilde{N}\ch + N^a\ch_a-\o_{0i}\cg ^i
+\l_{ab}(\tilde{N},\p,\o)\F^{ab}
\label{4.40} \ee
where $\l_{ab}$ is the specific function of $\tilde{N}, \p^{ai}$ and $\o_{ai}$,
given
in
(\ref{4.37}) and (\ref{4.39}).
There are twelve second-class constraints: $\F^{ab}$ and $\J^{ab}$, and ten
first
class constraints: $\cg ^i, \ch_a$ and $\tilde{H}:=\ch +\frac{1}{\tilde{N}}
\l_{ab}(\tilde{N},\p,\o)\F^{ab}$.
The number of degrees of freedom are: half the number of phase space
coordinates
(18) minus the number of first-class constraints (10) minus half the number of
second-class constraints (6), which gives two degrees of freedom per point.

The next task in this section is to eliminate the second-class constraints from
the
theory, leaving a Hamiltonian formulation with first-class constraints only. In
doing
this elimination, I find it convenient to switch back to the original $SO(1,3)$
notation. First, to solve $\F^{ab}\approx0$, I use the origin of that
constraint;
namely (\ref{4.3}). So, I fix a trio of non-degenerate
space-like $SO(1,3)$ vectors $E^{aI}$, and also the (up to a sign) unique
time-like unit vector, orthogonal to $E^{aI}$, $N^I$.

\bea det(E^{aI}E^b_I)&>&0 \nn \\
E^{aI}N_I&=&0 \nn \\
N^IN_I&=&-1 \label{4.41} \eea
$N^I$ can be explicitly defined as

\be N^I=-\frac{\e^{IJKL}E^a_JE^b_KE^c_L\e_{abc}}{6\sqrt{det(E^{aI}E^b_I)}}
\label{4.42}
\ee
Note the similarity between the pair $(N^I,E^{aJ})$ and $(N^I,V^{aJ})$ in
 (\ref{4.5}).
It is also useful to define the projection operator

\be \tilde{\eta}^{IJ}:=E^{aI}E_a^I=\eta^{IJ}+N^IN^J \label{4.43} \ee
where $E_a^I:=E_{ab}E^{bI}$, and $E_{ab}$ is the inverse to $E^{aI}E^b_I$.
Using this projection operator, one can separate all $SO(1,3)$ indices into
"boost"
and "rotational" parts. The solution to $\F^{ab}\approx0$ can now be written as

\be \p^{aIJ}=N^{[I}E^{aJ]} \label{4.44} \ee
where the twelve physical degrees of freedom in $\p ^{aIJ}$ now is captured in
$E^{aI}$. Now it is possible to solve also $\J^{ab}$ for $\o _{aIJ}$, but
instead of
doing so, I will take a shortcut via the symplectic form $\p
^{aIJ}\dot{\o}_{aIJ}$.
I
use the solution to $\F^{ab}\approx0$, (\ref{4.44}) in the symplectic form, and
see
what components of $\o_{aIJ}$, $E^{aI}$ will project out. To do that, I first
split $\o
_{aIJ}$ into two separate pieces

\be \o _{aIJ}=M_{aIJ}+\G_{aIJ} \label{4.45} \ee
where $\G_{aIJ}$ is the unique torsion-free connection compatible with
$E^{aI}$:

\be
D_aE^{bI}=\partial_aE^{bI}+\G^b_{ac}E^{cI}+\G_a{}^I{}_JE^{bJ}-\G^c_{ca}E^{bI}=0
\label{4.455} \ee
See Appendix B for a treatment of this type of "hybrid" spin-connection.
Thus, (\ref{4.45}) can be seen as a definition of $M_{aIJ}$. Then, in order to
rewrite
the symplectic form I need some formulas relating $\dot{N}^I$ and
$\dot{E}^{aI}$.
 From (\ref{4.41}) it follows that

\bea \dot{N}^IN_I&=&0 \nn \\
\dot{N}^IE^a_I&=&-\dot{E}^a_IN^I \label{4.465} \eea
Now, the symplectic form becomes

\bea
\p^{aIJ}\dot{\o}_{aIJ}&=&2N^IE^{aJ}(\dot{\G}_{aIJ}+\dot{M}_{aIJ})=2N^I\left
(\dot{D_aE^a_I}-
D_a\dot{E}^a_I+E^{aJ}\dot{M}_{aIJ}\right ) \nn \\
&&=-2\partial_a(N^I\dot{E}^a_I)+2N^IE^{aJ}\dot{M}_{aIJ}\nn \\
&&=-2\partial_a(N^I\dot{E}^a_I)+2\dot{(N^IE^{aJ}M_{aIJ})}
+\dot{E}^{aJ}(2N_JE_{a}^IE^{bK}M_{bIK}-2N^IM_{aIJ}) \nn \\ \label{4.46} \eea
Neglecting the surface terms, one can now read off the momenta conjugate to
$E^{aI}$,

\bea \{E^{aI}(x),K_{bJ}(y)\}&=&\d^a_b\d^I_J\d^3(x-y) \nn \\
K_{aJ}&:=&2N_JE_{a}^IE^{bK}M_{bIK}-2N^IM_{aIJ} \label{4.47} \eea
This means that the physical components $E^{aI}$ of $\p^{aIJ}$, project out
twelve
components $K_{aI}$ of $\o _{aIJ}$. The question is then; are these twelve
components
$K_{aI}$
of $\o _{aIJ}$, the twelve components that survives the constraint $\J^{ab}$?
To answer
this question, I first invert the relation (\ref{4.47}).

\bea
M_{aIJ}&=&-\frac{1}{2}E^{bK}K_{aK}N_{[J}E_{bI]}-
\frac{1}{4}N^KK_{eK}E^e_{[I}E_{aJ]}
\nn \\
&&+ T_a{}^{bc}E_{bI}E_{cJ} \label{4.48} \eea
where $T_a{}^{(bc)}=0$ and $T_a{}^{ab}=0$. $T_a{}^{bc}$ are the six components
of $\o
_{aIJ}$ that is orthogonal to $E^{aI}$ in the symplectic form. Now, putting
this into
the other second-class constraint $\J^{ab}$, one gets

\bea
\J^{ab}&=&-\frac{1}{2}\p^{cIJ}\e_{IJKL}\p^{(bLM}\cd_c\p^{a)}{}_M{}^K=\cdots
\sim
\e^{cd(a}T_c{}^{b)g}E_{gd} \label{4.50} \eea
And requiring that $\J^{ab}=0$ imposes six conditions on the six components in
$T_a{}^{bc}$, which can easily be solved to get: $T_a{}^{bc}=0$.
This means that the solution to $\F^{ab}=0$ really projects out the solution to
the
other
second-class constraint $\J^{ab}$, and that
the physical degrees of freedom surviving both
$\F^{ab}$ and $\J^{ab}$ are captured in $E^{aI}$ and $K_{aI}$. To summarize the
solution of the second-class constraints, we have

\bea \p^{aIJ}&=&N^{[I}E^{aJ]} \nn \\
\o_{aIJ}&=&\G_{aIJ}-\frac{1}{2}E^{bK}K_{aK}N_{[J}E_{bI]}-
\frac{1}{4}N^KK_{eK}E^e_{[I}E_{aJ]} \nn \\
\{E^{aI}(x),K_{bJ}(y)\}&=&\d^a_b\d^I_J\d^3(x-y) \label{4.505} \eea
With this solution of the second-class constraints at hand, it is time to
rewrite the
total Hamiltonian in terms of the physical fields $E^{aI}$ and $K_{aI}$. A
straightforward calculation gives

\bea \cg _{IJ}&=&\cd _a\p^{a}{}_{IJ}=\cdots=-\frac{1}{2}E^a_{[I}K_{aJ]}\approx0
\label{4.52}\\
\ch _a&=&\p^{bIJ}R_{abIJ}=\cdots=-D_{[a}(E^{bM}K_{b]M})\approx0 \label{4.53} \\
\ch &=&\p^{aIJ}\p^b{}_{JK}R_{ab}{}^K{}_I-2\l
\sqrt{det(\p^{aIJ}\p^b_{JI})}=\cdots
\nn \\
&=&-E^{aI}E^{bJ}\tilde{R}_{abIJ}-2\l \sqrt{det(E^{aI}E^b_I)}-\frac{1}{4}E^{aI}
E^{bJ}
K_{[aI}K_{b]J}\approx0 \label{4.54} \eea
where $D_a$ is the $E^{aI}$ compatible torsion-free covariant derivative, and
$\tilde{R}_{abIJ}$ is its curvature: $\tilde{R}_{abIJ}:=\partial_{[a}\G_{b]IJ}+
[\G_a,\G_b]_{IJ}$. In equations (\ref{4.53}) and (\ref{4.54}), I have neglected
terms
proportional to Gauss' law $\cg ^{IJ}$.

But this Hamiltonian is easily identified as
 the well-known ADM-Hamiltonian with the full $SO(1,3)$ invariance
unbroken. If one wants to compare this to section \ref{ADM} one first needs to
gauge-fix $E^{a0}=0$, and then solve the corresponding "boost-part" of Gauss'
law, $\cg
^{0i}=0$ $(K_{a0}=0)$. One should also de-densitize the coordinate $E^{ai}$. I
will
not do this comparison in detail, I just end this section by a short summary of
what has been done here.

The straightforward Hamiltonian formulation of the first-order H-P Lagrangian
gave a
Hamiltonian system with twelve second-class constraints. The reason why the
second-class constraints appear can be traced to the mismatch between the
number of
algebraically independents components of the spin-connection and the tetrad
field. (In
(2+1)-dimensions or in (3+1)-dimensions with only a self dual spin-connection,
the
mismatch disappears and there are no second-class constraints.) Then, when the
second-class constraints are solved to get a Hamiltonian containing only
first-class
constraints, one ends at the ADM-Hamiltonian.

In doing this Hamiltonian analysis of the full Hilbert-Palatini Lagrangian, it
is
striking how complicated the analysis is, compared to the analysis of the self
dual
H-P Lagrangian,
in next section. And the reason why the constraint analysis, in
this
section, is so complicated is of course the appearance of second-class
constraints.
We will see
that a reduction of the number of algebraically independent components in the
spin-connection will significantly simplify the Hamiltonian analysis. This
reduction
is accomplished by only using the self dual part of the spin-connection in the
Lagrangian.

\section{Self dual Hilbert-Palatini Lagrangian} \markboth{Chapter \thechapter \
\ \
Actions in (3+1)-dimensions}{\thesection \ \ \
 Self dual Hilbert-Palatini Lagrangian} \label{SDHP}

In this section, the H-P Lagrangian containing only the self dual part of the
curvature will be examined. I will first show that the equations of motion
following
from it, are the same as those from the full H-P Lagrangian. This is due to the
fact
that the two Lagrangians give the same equation of motion from the variation of
the
spin-connection, and once this equation is solved, the two Lagrangians differs
only by
a term that vanishes due to the Bianchi-identity.

After proving that this action is a good action for gravity, the Legendre
transform
will be performed, and the theory will be brought into the Ashtekar Hamiltonian
formulation. In
this Hamiltonian formulation, it will become clear that all the second-class
constraints from section \ref{HHP} now have disappeared. And the reason why
they do
not appear, is that with only the self dual spin-connection present, the
tetrad has enough independent components to function as momenta to the
spin-connection, without restrictions.

The analysis of the self dual H-P Lagrangian was first performed in \cite{9}.
 Later, this Lagrangian has been examined in \cite{7},
\cite{8}, \cite{10} and \cite{11}.

Before I start working on the action, I will give a few basic features of
self duality. Consider an $so(1,3)$ Lie-algebra valued field: $A^{IJ}$. The
dual
of $A^{IJ}$ is defined as

\be A^{*IJ}:=\frac{1}{2}\e^{IJ}{}_{KL}A^{KL} \label{5.1} \ee
and the dual of the dual becomes

\be A^{**IJ}=\frac{1}{4}\e^{IJ}{}_{KL}\e^{KL}{}_{MN}A^{MN}=-A^{IJ} \label{5.2}
\ee
The minus-sign follows from the Lorentz-signature of the Minkowski metric. (For
Euclidean signature, $so(4)$, there is a plus-sign there instead.) Now, since
the
duality-operation imposed twice has the eigen-value minus one for any $A^{IJ}$,
it is
possible to diagonalize $A^{IJ}$ as follows

\be A^{IJ}=A^{(+)IJ} + A^{(-)IJ} \label{5.3} \ee
where

\be A^{*IJ}=+i A^{(+)IJ} -i A^{(-)IJ} \label{5.4} \ee
$A^{(+)IJ}$ and $A^{(-)IJ}$ are called the self dual and the anti-self dual
part of
$A^{IJ}$. They can be explicitly defined as

\be A^{\pm IJ}:=\frac{1}{2}(A^{IJ}\mp i
A^{*IJ})=\frac{1}{2}(A^{IJ}\mp\frac{i}{2}\e^{IJ}{}_{KL}A^{KL}) \label{5.6} \ee
The self dual and anti-self dual parts can in some sense be seen as orthogonal
components, and the following relations are easily proven with (\ref{5.6}) and
the
$\e-\d$
identity.

\bea A^{(+)IJ}B^{(-)}_{IJ}&=&0 \label{5.7} \\
\Rightarrow\; A^{(+)IJ}B_{IJ}&=&A^{(+)IJ}B^{(+)}_{IJ} \label{5.8} \\
\left[A,B\right]^{IJ}&=&[A^{(+)},B^{(+)}]^{IJ}+[A^{(-)},B^{(-)}]^{IJ}
\label{5.9} \eea
Equation (\ref{5.7}) is just the orthogonality relation, and (\ref{5.9}) shows
that
the complexification of the Lorentz-algebra splits into its self dual and
anti-self dual sub-algebras: $so(1,3;C)\sim so(3)\times so(3)$. Equation
(\ref{5.9})
also shows that the self dual curvature is the curvature of the self dual
spin-connection:

\be R_{\a \b}^{IJ}(\o _\a ^{(+)KL})=R_{\a \b}^{(+)IJ}(\o _\a ^{(+)KL})=
R_{\a \b}^{(+)IJ}(\o _\a ^{KL}) \label{5.10} \ee
Now, since the original $A^{IJ}$ has 6 algebraically independent components,
while
$A^{(+)IJ}$ and $A^{(-)IJ}$ has only three algebraically independent
components, there
exist relations between different components of $A^{(+)IJ}$ and $A^{(-)IJ}$.
For instance

\bea A^{(\pm)}_{kl}&=&\pm i A^{(\pm)0i}\e_{ikl} \label{5.11} \\
A^{(\pm)0i}&=&\mp \frac{i}{2}\e^{ikl}A^{(\pm)}_{kl} \label{5.12} \eea
where $\e^{ijk}:=\e^{0IJK}$ and $\e_{ijk}=\e^0{}_{IJK}$. This split can also be
made
 Lorentz-covariant. (Note that, if $A^{(+)IJ}$ is a Lorentz-covariant object,
it
 transforms only under the self dual Lorentz-transformations, while a
self dual Lorentz-connection transforms under the full Lorentz-group.)

To make the split (\ref{5.11}) and (\ref{5.12}) Lorentz covariant, I define a
trio of
non-degenerate space-like $SO(1,3)$ vector-fields $V^{aI}$, and also the (up to
a
sign) unique time-like unit vector-field $N^I$, orthogonal to $V^{aI}$.

\be det(V^{aI}V^b_I)>0,\hspace{10mm}N_IV^{aI}=0,\hspace{10mm}N^IN_I=-1
\label{5.13} \ee
An explicit definition of $N^I$ is

\be N^I:=-\frac{\e^{IJKL}V^a_JV^b_KV^c_L\e_{abc}}{6\sqrt{det(V^{dM}V^e_M)}}
\label{5.14} \ee
I also define the projection operator

\be \tilde{\eta}^{IJ}:=V^{aI}V_a^J=\eta ^{IJ}+N^IN^J \label{5.15} \ee
where $V_a^I:=V_{ab}V^{bI}$, and $V_{ab}$ is the inverse to
$V^{ab}:=V^{aI}V^b_I$.
Using this projection operator, all $so(1,3)$ indices can be decomposed into
parts
parallel and orthogonal to $N^I$.

\bea K^{I'}&:=&-N^IN_JK^J \label{5.16} \\
K^{\tilde{I}}&:=&\tilde{\eta}^I_JK^J \label{5.17} \\
\Rightarrow\; K^I&=&K^{I'}+K^{\tilde{I}} \label{5.18} \eea
Note that

\be N^I=N^{I'},\hspace{10mm}V^{aI}=V^{a\tilde{I}},\hspace{10mm}
\tilde{\eta}^{IJ}=\tilde{\eta}^{\tilde{I}\tilde{J}} \label{5.19} \ee
I also define

\be \e^{\tilde{J}\tilde{K}\tilde{L}}:=N_I\e^{IJKL}
\Rightarrow\e^{I'\tilde{J}\tilde{K}
\tilde{L}}=-N^I\e^{\tilde{J}\tilde{K}\tilde{L}} \label{5.20} \ee
Then, using the $\e-\d$ identity for the full $\e^{IJKL}$, it is possible to
show that

\be \e^{\tilde{J}\tilde{K}\tilde{L}}\e_{\tilde{M}\tilde{N}\tilde{P}}=\d^{
[\tilde{J}
\tilde{K}\tilde{L}]}_{\tilde{M}\tilde{N}\tilde{P}} \label{5.21} \ee

Now, with all this machinery, it is straightforward to make the split
(\ref{5.11}),
(\ref{5.12}) covariant. First

\bea
A^{(+)I'\tilde{J}}&=&\frac{1}{2}(A^{I'\tilde{J}}+\frac{i}{2}N^I\e^{\tilde{J}
\tilde{K}\tilde{L}}A_{\tilde{K}\tilde{L}}) \label{5.22} \\
A^{(+)\tilde{I}\tilde{J}}&=&\frac{1}{2}(A^{\tilde{I}\tilde{J}}+iN^{K'}\e^{\tilde{I}
\tilde{J}\tilde{L}}A_{K'\tilde{L}}) \label{5.23} \eea
which give the relations:

\bea A^{(\pm)\tilde{I}\tilde{J}}&=&\pm i
N^{K'}A^{(\pm)}_{K'\tilde{M}}\e^{\tilde{I}\tilde{J}\tilde{M}} \label{5.24} \\
A^{(\pm)}_{K'\tilde{M}}&=&\pm\frac{i}{2}N_{K'}\e_{\tilde{I}\tilde{J}\tilde{M}}
A^{(\pm)
\tilde{I}\tilde{J}} \label{5.25} \eea
To compare (\ref{5.11}), (\ref{5.12}) with (\ref{5.24}), (\ref{5.25}), one can
make the
choice:\\ $N^I=(-1,0,0,0)$ $\Rightarrow$ $V^{a0}=0$.

Now, it is time to introduce the self dual Hilbert-Palatini Lagrangian.

\be \cl^{(+)}_{HP}:=e\left (\Sigma^{(+)\a \b}_{IJ}(e^{\g} _K)
R^{(+)IJ}_{\a \b}(\o ^{(+)KL})
+\l\right )
\label{5.26} \ee
where $\Sigma^{\a \b}_{IJ}(e^{\g} _K):=\frac{1}{2}e^\a _{[I}e^\b _{J]}$,
 and
$\Sigma^{(+)\a \b}_{IJ}
(e^{\g} _K)$ is its self dual part. $R^{(+)IJ}_{\a \b}$ is the self dual
part of the
curvature of the spin-connection, and as mentioned earlier, the self dual part
of the
curvature of the full spin-connection equals the full curvature of the self
dual part
of the spin-connection. See (\ref{5.10}).

First, when I want to show that the equations of motion following from
(\ref{5.26})
are the same as the ones following from the full H-P Lagrangian, I will regard
$R^{(+)IJ}_{\a \b}$ as the self dual part of the full curvature. That is

\be R^{(+)IJ}_{\a \b}(\o ^{(+)})=\frac{1}{2}\left (R^{IJ}_{\a
\b}(\o)-\frac{i}{2}\e^{IJ}{}_{KL}(R^{KL}_{\a
\b}(\o)\right ) \nn \ee
The equations of motion following from the variation of $\o _\a ^{IJ}$ are

\be \frac{\d S^{(+)}_{HP}}{\d \o _\b ^{IJ}}=-2\cd_\a (e\Sigma^{(+)\a
\b}_{IJ})=0
\label{5.27} \ee
But, since both $e^\a _I$ and $\o _\a ^{IJ}$ are real, the real part of
(\ref{5.27})
is just the normal zero-torsion condition. And the imaginary part of
(\ref{5.27}) is
 the dual of the real part, and therefore contains the same
information. Then, if one solves (\ref{5.27}) for $\o _\a ^{IJ}$, and uses that
solution in the Lagrangian (\ref{5.26}), the imaginary part of the Lagrangian
vanishes
due to the Bianchi-identity. And the real part of the Lagrangian is just the
conventional Hilbert-Palatini Lagrangian (or Einstein-Hilbert Lagrangian, when
the
solution to (\ref{5.27}) is used). So, altogether this means that the
variation of $\o _\a ^{IJ}$ implies the normal zero-torsion condition, and when
using
the solution to that equation, the variation of $e^\a _I$ implies Einstein's
equations.
Note that the reality condition on the spin-connection really is superfluous.
It is
enough to require the tetrad to be real, then equation (\ref{5.27}) will take
care
of the reality of the spin-connection. It is this fact that Ashtekar uses when
he only
imposes reality conditions on the metric-variables, and not on the connection.
Note,
however, that this is only true for non-degenerate metrics. With a degenerate
metric,
(\ref{5.27}) cannot be solved to get a unique spin-connection.

Before doing the (3+1)-decomposition and Legendre transformation of
(\ref{5.26}), I
will eliminate the non-independent parts of $\o ^{(+)IJ}$. Using (\ref{5.24})
and
(\ref{5.25}), it follows that

\be
A^{(+)IJ}B^{(+)}_{IJ}=2A^{(+)I'\tilde{J}}B^{(+)}_{I'\tilde{J}}+
A^{(+)\tilde{I}\tilde{J}} B^{(+)}_{\tilde{I}\tilde{J}}=4A^{(+)I'\tilde{J}}
B^{(+)}_{I'\tilde{J}} \label{5.28} \ee
This means that the (3+1)-decomposed Lagrangian becomes

\be \cl ^{(+)}_{HP}=e\left (8\Sigma^{(+)0 b}_{I'\tilde{J}}R^{(+)I'\tilde{J}}_{0
b}+
4 \Sigma^{(+)ab}_{I'\tilde{J}}
R^{(+)I'\tilde{J}}_{ab} +\l\right ) \label{5.29} \ee
Then, I define the momenta conjugate to $\o ^{(+)I'\tilde{J}}_b$:

\be \p ^b_{I'\tilde{J}}:=\frac{\d L}{\d \dot{\o}_b^{(+)I'\tilde{J}}}=8 e
\Sigma^{(+)0 b}_{I'\tilde{J}} \label{5.30} \ee
Remember that $\Sigma^{\a \b}_{IJ}$ is not an independent field, it is just the
anti-symmetric product of two tetrads. The next step in the Legendre transform
is to
rewrite the Lagrangian in terms of the phase space variables, but before doing
so, I
will introduce the ADM-like tetrad decomposition. I decompose the tetrad as
follows

\bea e^{0I}&=&-\frac{N^I}{N} \hspace{10mm}\Rightarrow \hspace{10mm}
e^{0I}e^0_I=-\frac{1}{N^2} \nn \\
e^{aI}&=&V^{aI} + \frac{N^a\;N^I}{N};\hspace{10mm}V^{aI}N_I=0 \label{5.31} \eea
This decomposition is not a restriction on the tetrad, the tetrad is completely
general. Instead, it is a choice for the previously introduced vector fields:
$N^I,\;V^{aI}$. Note that this decomposition gives the standard ADM-form of the
metric:

\be g^{\a \b}=\left(\begin{array}{cc}
-\frac{1}{N^2}& \frac{N^a}{N^2}\\
\frac{N^a}{N^2}& V^{aI}V^b _I -\frac{N^aN^b}{N^2}
\end{array} \right) \nn \ee
Using this decomposition in (\ref{5.30}), gives

\be \p ^b_{I'\tilde{J}}=2e(e^{[0}_{I'}e^{b]}_{\tilde{J}}-i
\e_{I'\tilde{J}}{}^{\tilde{K}\tilde{L}}
e^0_{\tilde{K}}e^b_{\tilde{L}})=-\frac{2e}{N}N_IV^b_J \label{5.32} \ee
Note that the imaginary part of $\p ^a _{I'\tilde{J}}$ automatically vanishes
with
this choice of unit time-like vector-field. Here, I want to emphasize that it
is not
a gauge-choice involved here, the tetrad is completely general, the choice lies
instead in what twelve components of $\o ^{(+)IJ}_\a$ that should be regarded
as
independent, and the clever choice is to choose $\o ^{(+)IJ}_\a$'s projection
along
$e^{0I}$. This choice is clever, since it makes the momenta real and simple.
Note,
however, that there is nothing that says that the momenta must be real. A
priori, we
only know that the tetrad is real, and that can be achieved by the weaker
requirement
 $Im(\p ^a_{I'\tilde{J}}\p ^{bI'\tilde{J}})=0$.

Now, returning to the Legendre transform, relation (\ref{5.32}) should be
inverted,

\be V^b_J=\frac{N}{2e}N^I\p ^b_{IJ} \label{5.33} \ee
and the two last terms in (\ref{5.29}) should be rewritten in terms of the
momenta.

\bea 4e \Sigma^{(+)ab}_{I'\tilde{J}}R^{(+)I'\tilde{J}}_{ab} &=& -N^a\p
^b_{I'\tilde{J}}
R^{(+)I'\tilde{J}}_{ab}-\frac{i N^2}{4e}\p ^a_{I'\tilde{K}}\p ^{bI'}{}
_{\tilde{L}} N^{N'}
R^{(+)}_{abN'\tilde{M}}\e^{\tilde{K}\tilde{L}\tilde{M}} \nn \\
e &=&\frac{N^2}{8e} det\p ^a_{I'\tilde{J}}=\frac{N^2}{8e} \frac{1}{6}\e_{abc}
\e^{\tilde{J}\tilde{K}\tilde{L}}\p^a_{M'\tilde{J}}\p^b_{N'\tilde{K}}\p^c_{P'\tilde{L}}
N^{M'}N^{N'}N^{P'} \nn \\ \label{5.35} \eea
Before I write out the total phase space Lagrangian, I introduce the Ashtekar
variables:

\bea A_{\a i}&:=&-2 N^J\o ^{(+)}_{\a J'\tilde{I}} ,\hspace{5mm}F_{abi}:=-2
N^{J'}R^{(+)}
_{abJ'\tilde{I}}=\partial_{[a}A_{b]i}+i \e_{ijk}A_a^jA_b^k \nn \\
E^{ai}&:=&\frac{1}{2}N_{J'}\p
^{aJ'\tilde{I}},\hspace{5mm}\tilde{N}:=\frac{N^2}{e}
\label{5.37} \eea
where I used the fact that, since the "tilded" indices are orthogonal to a
time-like
direction, they are really $SO(3)$ indices. ($i,\;j,\;k\;...$ are $SO(3)$
indices.) In
terms of these variables, the total Lagrangian becomes

\be \cl ^{(+)}_{HP}=E^{ai}\dot{A}_{ai}-N^a\ch _a-\tilde{N}\ch +A_{oi}\cg ^i
\label{5.38} \ee
where

\bea \ch _a&:=&E^{bi}F_{abi}\approx0 \nn \\
\ch &:=&\frac{i}{2}E^{ai}E^{bj}F_{ab}{}^k\e_{ijk}-\l det(E^{ai})\approx 0 \nn
\\
\cg ^i &:=&\cd _aE^{ai}=\partial _aE^{ai}+i \e^{ijk}A_{aj}E^a_k \approx0 \nn
\eea
are the constraints that follow from the variation of the Lagrange multiplier
fields
$N^a,\; \tilde{N},$ and $ A_{0i}$. The fundamental Poisson bracket is
$\{A_{ai}(x),E^{bj}(y)\}=\d ^b_a\d ^j_i\d ^3(x-y)$.

This is the famous Ashtekar formulation, and to
complete the
analysis of it, a constraint analysis must be performed. This will be done in
the next
section. But, before I leave this section, I will give the metric formula in
terms
of Ashtekar's variables.
Using (\ref{5.31}) and (\ref{5.37}), the densitized spacetime metric is

\be \tilde{g}^{\a \b}=\sqrt{-g}g^{\a \b}=\left(\begin{array}{cc}
-\frac{1}{\tilde{N}}& \frac{N^a}{\tilde{N}}\\
\frac{N^a}{\tilde{N}}& \tilde{N}E^{ai}E^b _i -\frac{N^aN^b}{\tilde{N}}
\end{array} \right) \label{5.41} \ee

The most appealing feature of the Ashtekar formulation is perhaps the simple
and
polynomial form of all the constraints. It is this feature that has enkindled
the old
attempts of non-perturbative canonical quantization of gravity. Another fact
that is
new here in the Ashtekar formulation, compared to the other Hamiltonians in
section
\ref{ADM} and \ref{HHP}, is that the phase space coordinate here is a
(gauge-)connection. In the ADM-Hamiltonian, the phase space variables are all
gauge-covariant objects, and in the Hamiltonian formulation of the full H-P
Lagrangian, the spin-connection can only be used as the phase space coordinate
if one
is prepared to pay the price of keeping second-class constraints in the theory.
Once
these constraints are eliminated, the coordinate is again a gauge-covariant
object.
This fact, that the phase space coordinate is a gauge-connection has made it
feasible
to import techniques and methods from the more well-studied analysis of
Yang-Mills
theories.

\section{The Ashtekar Hamiltonian} \markboth{Chapter \thechapter \ \ \
Actions in (3+1)-dimensions}{\thesection  \ \  \ The Ashtekar Hamiltonian}
\label{AH}

Here, I will do the constraint analysis for the Ashtekar Hamiltonian given in
section
(\ref{SDHP}). Then, the reality conditions will be examined in more detail, and
finally
the canonical transformation, relating this formulation to the ADM-Hamiltonian,
will
be given.

The Ashtekar Hamiltonian was originally found through the above mentioned
canonical
transformation \cite{2}, and it soon became clear that this Hamiltonian
could be very useful in attempts to quantize gravity canonically. Prior to the
existence of this Hamiltonian, the attempts to quantize gravity canonically had
all
started from the complicated ADM-Hamiltonian. The ADM-Hamiltonian is
complicated
mostly due to the non-polynomial and inhomogeneous form of the
Hamiltonian constraint. This complicated form has made it practically
impossible to
find any quantum solution, in this formulation. Ashtekar's Hamiltonian,
however, has a
simple polynomial and homogeneous Hamiltonian constraint, and quantum solutions
to
this formulation were soon found \cite{3a}, \cite{3b}.

The Ashtekar Hamiltonian formulation can be summarized as:

\bea \ch _{tot} &:=&\tilde{N}\ch + N^a\ch _a -A_{0i}\cg ^i \label{6.1} \\
\ch _a&:=&E^{bi}F_{abi}\approx0 \nn \\
\ch &:=&\frac{i}{2}E^{ai}E^{bj}F_{ab}{}^k\e_{ijk}-\l det(E^{ai})\approx 0 \nn
\\
\cg ^i &:=&\cd _aE^{ai}=\partial _aE^{ai}+i \e^{ijk}A_{aj}E^a_k \approx0 \nn
\eea
The fundamental Poisson bracket is $\{A_{ai}(x),E^{bj}(y)\}=\d ^b_a\d ^j_i\d
^3(x-y)$,
and the other fields are Lagrange multiplier fields, whose variation implies
the
constraints given in (\ref{6.1}).
The densitized spacetime metric can be constructed from the phase space fields,
in a
solution, as

\be \tilde{g}^{\a \b}=\sqrt{-g}g^{\a \b}=\left(\begin{array}{cc}
-\frac{1}{\tilde{N}}& \frac{N^a}{\tilde{N}}\\
\frac{N^a}{\tilde{N}}& \tilde{N}E^{ai}E^b _i -\frac{N^aN^b}{\tilde{N}}
\end{array} \right) \label{6.105} \ee

And the reality conditions will be given later.
Next, I will show that this is a consistent Hamiltonian formulation in the
sense that
a field-configuration that initially satisfies the constraints will stay on the
constraint surface under time evolution. This is the same as requiring the
time evolution of the constraints to be weakly vanishing. To show that, I will
calculate the constraint algebra. First, the transformations generated by $\cg
^i$ are
given:

\bea \d ^{\cg ^i}E ^{ai}&:=&\{E ^{ai},\cg ^j[\L_j]\}=i\e^i{}_{jk}\L^jE ^{ak}
\label{6.3} \\
\d ^{\cg ^i}A _a ^i&=&\{A _a ^i,\cg ^j[\L_j]\}=-\cd _a\L^i \label{6.4} \eea
which shows that $\cg ^i$ is the generator of $SO(3)$ transformations. (Note
that
$SO(3)$ here is the self dual part of the Lorentz-group, not the rotation
part.) Then,
I define the generator of spatial diffeomorphisms:

\be \tilde{\ch}_a:=\ch _a-A_{ai}\cg ^i \label{6.5} \ee
and calculate the transformations it generates:

\bea \d ^{\tilde{\ch}_a}E ^{ai}&:=&\{E
^{ai},\tilde{\ch}_b[N^b]\}\!=\!N^b\partial _b
 E
^{ai} \!-\! E ^{bi}\partial _b N^a \!+\! E ^{ai} \partial _b N^b\!=\!
\pounds _{N^b}E ^{ai}
\label{6.6} \\
\d ^{\tilde{\ch}_a}A _a ^i&=&\{A _a ^i,\tilde{\ch}_b[N^b]\}\!=\!N^b\partial _b
A _a ^i \!+\!
A _{b}^i\partial _a N^b\!=\!\pounds _{N^b}A _a ^i \label{6.7} \eea
Now, one needs to know how the constraints transform under $SO(3)$ rotations
and
spatial diffeomorphisms. Under $SO(3)$ transformations; $\cg ^i$ is a vector,
$\tilde{\ch}_a$ is non-covariant, and $\ch$ is a scalar. Under spatial
diffeomorphisms; $\cg ^i$ is a scalar density of weight plus one,
$\tilde{\ch}_a$ is a
covariant vector density of weight plus one, and $\ch$ is a scalar density of
weight
plus two. Note that, although $\tilde{\ch}_a$ is not covariant under $SO(3)$
rotations, the Poisson bracket between $\cg ^i$ and $\tilde{\ch}_a$ can yet
easily be
calculated since $\cg ^i$ transforms covariantly under spatial diffeomorphisms.

\bea \{\cg ^i[\L _i],\cg ^j[\g _j]\}&=&\cg _i[\L ^k\g ^j \e_{kj}{}^i]
\label{6.8} \\
\{\ch [N],\cg ^j[\g _j]\}&=&0 \label{6.9} \\
\{\cg ^i[\L _i],\tilde{\ch}_a[N^a]\}&=&\cg _i[-\pounds _{N^a}\L ^i]
\label{6.10} \\
\{\tilde{\ch}_b[M^b],\tilde{\ch}_a[N^a]\}&=&\tilde{\ch}_b[-\pounds _{N^a}M^b]
\label{6.11} \\
\{\ch [N],\tilde{\ch}_a[N^a]\}&=&\ch [-\pounds _{N^a}N] \label{6.12}  \eea
The only Poisson bracket left to calculate is $\{\ch ,\ch \}$, and again this
calculation simplifies by noting that only terms with derivatives on either $N$
or $M$
contribute:

\be \{\ch [N],\ch [M]\}=\ch _a[E^{ai}E^b_i(N\partial_b M-M\partial_b N)]
\label{6.13}
\ee
This shows that all the constraints are first-class, and since the total
Hamiltonian
is a linear combination of these constraints, the time evolution of the
constraints are
automatically consistent.

In section \ref{SDHP} it was shown that it is sufficient to require the tetrad
to be
real in order to get real general relativity from the self dual H-P Lagrangian.
With a
real tetrad, the spin-connection will automatically be real in a solution to
the
equations of motion. How can this be translated into the Hamiltonian
formulation, where
one normally wants to impose all restrictions on the phase space variables? The
equation that took care of the reality condition for the spin-connection was:
$\frac{\d S^{(+)}_{HP}}{\d \o ^{IJ}_\a}=0$, and here this becomes:

\bea \frac{\d S^{(+)}_{HP}}{\d A_{0i}}&=&\cg ^i=0 \label{6.01} \\
\frac{\d S^{(+)}_{HP}}{\d A_{ai}}&=&\dot{E}^{ai}+\frac{\d H_{tot}}{\d A_{ai}}=0
\label{6.02} \eea
In section \ref{SDHP} it was enough to require the tetrad to be real, and the
solution to (\ref{6.01}), (\ref{6.02}) gave automatically a real
spin-connection.
(Note, however, that $A_{ai}$, which in a solution is the self dual part of
this real
spin-connection, is {\em not} real.)
Here, in the Hamiltonian formulation, this is a bit unconvenient since the
velocity
$\dot{E}^{ai}$ appears in (\ref{6.02}), and requiring that $\dot{E}^{ai}$ is
real will
lead to conditions on the Lagrange multiplier fields $\tilde{N}$ and $A_{0i}$.
Also,
requiring the momenta $E^{ai}$ to be real will forbid complex $SO(3)$
gauge-transformations, which is a bit unnatural since $SO(3)$ here really
stands for
the self dual part of $SO(1,3)$, which is a complex Lie-algebra.

The way out of this is to concentrate the reality requirements on the metric
instead
of the tetrad.

\be Im(\tilde{N})=0,\hspace{10mm}Im(N^a)=0,\hspace{10mm}Im(E^{ai}E^b_i)=0
\label{6.04}
\ee
Then, imposing that $\dot{(E^{ai}E^b_i)}$ should be real, will, through its
equation of
motion, lead to:

\be Im(E^{cj}E^{(ai}\cd _cE^{b)k}\e_{ijk})=0 \label{6.05} \ee
which is a good condition in the sense that it does not involve any
Lagrange multipliers. Using the reality conditions (\ref{6.04}) and
(\ref{6.05}) it is
then straightforward to show that the metric will stay real under time
evolution,
provided the metric is invertible.\\ \\

Now, I want to show the relation between this Ashtekar Hamiltonian and the
conventional ADM-Hamiltonian in triad form. They are related through a complex
canonical transformation, and to show that, I first define the new field
$K_{ai}$:

\be K_{ai}:=A_{ai}+i\G _{ai}(E) \label{6.14} \ee
where $\G_{ai}$ is the unique torsion-free spin-connection defined to
annihilate
$E^{ai}$:

\be
D_aE^{bi}:=\partial_aE^{bi}+\G^b_{ac}E^{ci}-\G^c_{ac}E^{bi}+\e^{ijk}\G_{aj}E^{b}_k=0
\label{6.15} \ee
Now, using the definition (\ref{6.14}) to rewrite all the constraints in terms
of
$K_{ai}$ and $E^{ai}$, one gets

\bea \cg ^i&=&\cd_aE^{ai}=D_aE^{ai}+i\e^{ijk}K_{aj}E^a_k = i\e^{ijk}K_{aj}E^a_k
 \label{6.16} \\
\ch _a&=&E^{bi}(-i R_{abi}+D_{[a}K_{b]i}+i\e_{ijk}K_a^jK_b^k) \approx E^{bi}
D_{[a}K_{b]i} \label{6.17} \\
\ch&=&\frac{1}{2}E^{ai}E^{bj}R_{ab}^k\e_{ijk}-\l det(E^{ai})
+\frac{i}{2}E^{ai}E^{bj}D_{[a}K_{b]}^k\e_{ijk}
-\frac{1}{2}E^{ai}E^{bj}K_{[ai}K_{b]j}
\nn \\
&\approx&\frac{1}{2}E^{ai}E^{bj}R_{ab}^k\e_{ijk}-\l det(E^{ai})
-\frac{1}{2}E^{ai}E^{bj}
K_{[ai}K_{b]j} \label{6.18} \eea
where $R_{ab}^k:=\partial_{[a}\G_{b]}^k+\e^{klm}\G_{al}\G_{bm}$, and I have
used the
Bianchi-identity in (\ref{6.17}), and neglected terms proportional to $\cg ^i$
in both
(\ref{6.17}) and (\ref{6.18}). Comparing these constraints with the
ADM-constraints in
section \ref{ADM}, one notices that these constraints have exactly the same
structure. To get exact agreement, one needs to change $E^{ai}$ into its
non-densitized inverse $e_{ai}$, which is the coordinate used in section
\ref{ADM}. I
will not do this canonical transformation here, instead I will now show that
the
transformation (\ref{6.14}) really is a canonical one. To show that, I first
define
the undensized $E^{ai}$ and its inverse

\be e^{ai}:=\frac{1}{\sqrt{det(E^{bj})}}E^{ai},\hspace{10mm}e_{bi}e^{ai}=\d
^a_b
,\hspace{10mm}\tre e:=det e_{ai}=\sqrt{det E^{ai}}
\label{6.20} \ee
Then, the following is true

\be D_{[a}e_{b]i}=\partial_{[a}e_{b]i}+\e_{ijk}\G_{[a}^je_{b]}^k=0 \label{6.21}
\ee
and the time-derivative of (\ref{6.21}) must also vanish.

\be D_{[a}\dot{e}_{b]i}+\e_{ijk}\dot{\G}_{[a}^je_{b]}^k=0 \label{6.22} \ee
Solving (\ref{6.22}) for $\dot{\G}_a^i$, and contracting its indices with an
$E^{ai}$
then gives:

\be \dot{\G}_a^iE^a_i=-\frac{1}{2}\tre e e^{aj}e^{bk}\e_{ijk}D_a\dot{e}_b^i=-
\frac{1}{2}\partial_a(\tre e e^{aj}e^{bk}\e_{ijk}\dot{e}_b^i) \label{6.23} \ee
This means that $E^{ai}\dot{A}_{ai}=E^{ai}\dot{K}_{ai}$ up to the surface term
(\ref{6.23}), which shows that the fundamental Poisson bracket now is

\be \{K_{ai}(x),E^{bj}(y)\}=\d^b_a\d^j_i\d^3(x-y) \label{6.235} \ee
and that the transformation really is a canonical one. Note that a comparison
between
(\ref{6.18}) and the ADM-Hamiltonian in section \ref{ADM} shows that
$K_{ai}E^i_b$
has the interpretation as the (densitized) extrinsic curvature. Further details
about
this canonical transformation can be found in {\em e.g} \cite{2} and \cite{8b}.

\section{The CDJ-Lagrangian} \label{CDJ} \markboth{Chapter \thechapter \ \ \
Actions in (3+1)-dimensions}{\thesection \ \  \ The CDJ-Lagrangian}

 From section \ref{EH} to \ref{AH} the emphasis has gradually shifted from the
metric towards the connection as the fundamental field for gravity. Here in
this
section, the shift will reach almost completion, when the metric is eliminated
from
the Ashtekar Hamiltonian (except for the conformal factor), leaving the
connection as
the fundamental field.

The pure spin-connection formulation of gravity was first found \cite{12}
through a field-elimination from the Plebanski action in section
(\ref{PL}). It was, however, also clear that the Hamiltonian formulation of the
pure
spin-connection Lagrangian equals the Ashtekar Hamiltonian. So, the Ashtekar
Hamiltonian and the CDJ-Lagrangian are related via a Legendre transform.

In this section, I want to explicitly perform the Legendre transform, starting
from
the Ashtekar Hamiltonian, and in order not to complicate things more than
necessary, I
will put $\l =0$ here. The Legendre transform for $\l \neq 0$ is much more
complicated
to perform, and details can be found in
\cite{13}.\\ \\

 In doing the Legendre transform, I will also assume that the vector
constraint, in Ashtekar's Hamiltonian, can be found as a primary constraint to
the
resulting Lagrangian. This means that the equations that should be solved are

\bea
\dot{A}_{ai}&=&\frac{i\tilde{N}}{2}\e_{abc}\e_{ijk}E^{bj}B^{ck}+\frac{1}{2}N^b\e_{bac}
B^c_i +\cd _aA_{0i} \label{7.1} \\
\e_{abc}E^{bi}B^c_i&=&0 \label{7.2} \eea
where $B^{ai}$ is the magnetic field: $B^{ai}:=\e^{abc}F_{bc}^i$. Now, doing
the
Legendre transform means solving the equation of motion for the momentum, to
get
the momenta
as a function of the coordinate and its velocity. Here, this corresponds to
solving
(\ref{7.1}) for $E^{ai}$. Equation (\ref{7.2}) should then also be solved for
$E^{ai}$
in order to find $\ch _a$ as a primary constraint.

To be able to solve these two equations, I must require the magnetic field to
be
non-degenerate, i.e

\be det(B^{ai}):=\frac{1}{6}\e_{abc}\e_{ijk}B^{ai}B^{bj}B^{ck}\neq 0
\label{7.3} \ee
This restriction of the magnetic field will exclude some field-configurations.
In
\cite{12}, these excluded field-configurations are given in terms of
Petrov-classes. Note that flat Minkowski-space is excluded.
Now, if $B^{ai}$ is non-degenerate, it can serve as a basis for all spatial
vector-fields (or $SO(3)$ vector-fields). For instance

\be E^{ai}=\Psi^{ij}B^a_j \label{7.35} \ee
is then always true for some $SO(3)$ matrix $\Psi^{ij}$. Using this relation in
(\ref{7.2}), implies

\be \Psi^{[ij]}=0 \label{7.6} \ee
Then, multiplying (\ref{7.1}) by $B^{aj}$ and symmetrizing in the $SO(3)$
indices,
yields

\be \O^{ij}=\frac{1}{\eta}(\Psi ^{ij}-\d ^{ij}Tr\Psi) \label{7.8} \ee
where $\O^{ij}:=\e^{\a \b \g \d}F_{\a \b}^iF_{\g \d}^j=2B^{a(i}F_{oa}^{j)}$ and
$\eta:=-\frac{1}{2i\tilde{N}det(B^{ai})}$. This equation is easily solved

\be \Psi ^{ij}=\eta (\O^{ij}-\frac{1}{2}\d ^{ij}Tr\O) \label{7.9} \ee
and the Legendre transform is done. Putting this solution into the Lagrangian
$\cl=E^{ai}\dot{A}_{ai}-\ch^{Ash}$, one gets

\be \cl _{CDJ}=\frac{\eta}{8}(Tr\O^2-\frac{1}{2}(Tr\O)^2) \label{7.10} \ee
which is the sought for, pure spin-connection Lagrangian for pure gravity. Note
that
$\cl _{CDJ}$ is not a totally metric-free Lagrangian since $\eta$ is related to
the metric via
the lapse function. With a cosmological constant and/or matter fields it is,
however, in
principle possible to eliminate $\eta$ also. See \cite{13}. To get a formula
for
the metric here in the CDJ-formulation, one can return to the metric-formula in
the
Ashtekar formulation, and then follow the fields through the Legendre
transform. The
result is

\be \tilde{g}^{\a \b}=\sqrt{-g} g^{\a \b}=-\frac{2i}{3} \eta \e_{ijk}\e^{\a \g
\d
\e}F_{\g \d}^iF_{\e \k}^jF_{\m \n}^k\e^{\b \m \n \k} \label{7.11} \ee
which is called the Urbantke formula. Before the CDJ-formulation was known,
Urbantke
\cite{14}
gave the metric-formula (\ref{7.11}) as the solution to the problem: with
respect to
what metric is a given $SO(3)$ field strength, self dual? Since this question
is
insensitive to the conformal factor of the metric, Urbantke could only give the
metric
up to an arbitrary conformal factor, and the result was (\ref{7.11}). Further
details
about the CDJ-formulation can be found in: the original discovery \cite{12} and
\cite{12b}, relation to the Ashtekar Hamiltonian \cite{15}, \cite{16},
\cite{13} and
\cite{18}, with new cosmological
constants \cite{19}, \cite{17}, \cite{16b} and \cite{18},
generalization to other gauge groups, \cite{20}.

\section{The Plebanski Lagrangian} \markboth{Chapter \thechapter \ \ \
Actions in (3+1)-dimensions}{\thesection \ \  \ The Plebanski Lagrangian}
\label{PL}

In section \ref{SDHP}, we learned that it is enough to have only the self dual
part of the H-P Lagrangian in order to get full general relativity. This means
that the
only combination of the tetrad fields that is needed is the self dual part of
the
anti-symmetric product of two tetrads. Plebanski made use of that fact, and
promoted
this combination of tetrads into an independent field. Then, to still get the
same
physical content in the action, he had to add a Lagrange multiplier term,
imposing the
original relation.

Here, I want to briefly go through the steps that take us from the self dual
H-P
Lagrangian to the Plebanski Lagrangian. Then, I will also derive the
CDJ-Lagrangian by
eliminating fields from the Plebanski Lagrangian. Details about the Plebanski
action
can be found in \cite{21}, \cite{12}.

The self dual H-P Lagrangian:

\be \cl^{(+)}_{HP}:=e\left (\Sigma^{(+)\a \b}_{IJ}(e^{\g} _K)
R^{(+)IJ}_{\a \b}(\o ^{(+)KL})
+\l\right )
\label{8.1} \ee
where $\Sigma^{\a \b}_{IJ}(e^{\g} _K):=\frac{1}{2}e^\a _{[I}e^\b _{J]}$,
 and
$\Sigma^{(+)\a \b}_{IJ}(e^{\g} _K)$ is its self dual part.
Now, using the identity

\be e\frac{1}{2}e^\a _{[I}e^\b _{J]} =\frac{1}{4}\e^{\a \b \g \d}e_{\g}^Ke_\d
^L
\e_{IJKL} \label{8.2} \ee
and defining

\be \Sigma_{\a \b}^{IJ}:=\frac{1}{2}e_\a^{[I}e_\b^{J]} \label{8.3} \ee
it is straightforward to show that

\be e\Sigma^{(+)\a \b}_{IJ}=\frac{i}{2}\e^{\a \b \g \d}\Sigma^{(+)}_{\g \d IJ}
\label{8.5} \ee
The determinant of $e_\a^I$ can also be written solely in terms of $\Sigma$'s:

\be e:=dete_\a ^I=\frac{i}{6}\e^{\a \b \g \d} \Sigma^{(+)IJ}_{\a \b}
\Sigma^{(+)}_{\g \d IJ} \label{8.6} \ee
With all this, the Lagrangian now becomes

\be \cl^{(+)}_{HP}=\frac{i}{2} \e^{\a \b \g \d} \Sigma^{(+)IJ}_{\a
\b}\left (R^{(+)}_{\g
\d IJ} +\frac{\l}{3}\Sigma^{(+)}_{\g \d IJ}\right ) \label{8.7} \ee
And, since both $\Sigma$ and $R$ are self dual, they have only 18 algebraically
independent components. To make this more manifest, I do as in section
\ref{SDHP},
and split $\Sigma$ and $R$ into theirs boost and rotation parts:

\be A^{IJ}=A^{[I'\tilde{J}]}+A^{\tilde{I}\tilde{J}} \nn \ee
where $A^{IJ}$ is a general $so(1,3)$ valued field, and the primed index stands
for
the projection along a time-like unit vector-field $N^I$, while the tilded
index stands
 for
the projection into the orthogonal space-like surface orthogonal to $N^I$. See
section ({\ref{SDHP}) for details. For a self dual field $A^{IJ}$ it is then
true that

\be A^{IJ}B_{IJ}=2A^{I'\tilde{J}}B_{I'\tilde{J}} +A^{\tilde{I}\tilde{J}} B_
{\tilde{I}\tilde{J}}=4A^{I'\tilde{J}}B_{I'\tilde{J}} \nn \ee
Then, I introduce the $SO(3)$ notation:

\bea \Sigma^i_{\a \b}:=\Sigma^{(+)J'\tilde{I}}_{\a \b} N_{J'} \nn \\
F_{\a \b i}:=-2 R^{(+)}_{\a \b J'\tilde{I}} N^{J'} \nn \\
A_{ai}:=-2\o ^{(+)}_{J'\tilde{I}}N^{J'} \label{8.85} \eea
where the factor of two is introduced just to get agreement with the Ashtekar
curvature in section \ref{AH}. Now, the Lagrangian is

\be \cl^{(+)}_{HP}=i\e^{\a \b \g \d} \Sigma^i_{\a \b}\left (F_{\g \d
i}-\frac{2\l}{3}
\Sigma_{\g \d i}\right ) \label{8.9} \ee
Remember that $\Sigma^i_{\a \b}$ is still just a given function of the
tetrad-field.
The idea is now to let $\Sigma^i_{\a \b}$ become an independent field, and at
the
same time add a Lagrange multiplier term, imposing the original definition of
it. A
constraint that does the job, is

\bea \tilde{M}^{ij}&:=&M^{ij}-\frac{1}{3}\d^{ij}M^k_k=0 \nn \\
M^{ij}&:=&\e^{\a \b \g \d}\Sigma^i_{\a \b}\Sigma^j_{\g \d} \label{8.10} \eea
To show that this is a necessary condition on $\Sigma$, one can just put in the
original definition of $\Sigma ^i_{\a \b}$ in $M^{ij}$:

\be M^{IJKL}:=\e^{\a \b \g \d}\Sigma^{(+)IJ}_{\a \b} \Sigma^{(+)KL}_{\g
\d}=\frac{1}{2} e \e^{IJKL} +i\frac{e}{2}\eta^{I[K}\eta^{L]J} \nn \ee
which means that

\be
M^{ij}:=N_NN_KM^{NIKJ}=-i\frac{e}{2}\tilde{\eta}^{\tilde{I}\tilde{J}}=\frac{1}{3}
\tilde{\eta}^{\tilde{I}\tilde{J}} N_NN_K\tilde{\eta}_{\tilde{L}\tilde{M}}
M^{NLKM}=\frac{1}{3}\d^{ij}M^k_k \label{8.12} \ee
To get an indication that $\tilde{M}^{ij}=0$ also is a sufficient condition,
one may
count the number of degrees of freedom in $\Sigma$ constrained by
$\tilde{M}^{ij}$, and
compare with the degrees of freedom in $e_\a ^I$: $\Sigma$ has 18 independent
components, and is constrained by the traceless symmetric three-by-three matrix
$\tilde{M}^{ij}$, leaving 13 degrees of freedom. $e_\a ^I$ has 16 free
components but
the self dual anti-symmetric product is invariant under anti-self dual Lorentz
transformations, leaving again 13 degrees of freedom. An explicit proof of the
adequacy of $\tilde{M}^{ij}=0$ can be found in \cite{12}. Now, when adding the
constraint $\tilde{M}^{ij}=0$ to the Lagrangian with a Lagrange multiplier, the
tracelessness of $\tilde{M}^{ij}$ can be shifted over to the Lagrange
multiplier, and
then I let that condition also be implied by a Lagrange multiplier term,

\be \cl=i\e^{\a \b \g \d}\Sigma^i_{\a \b}\left (F_{\g \d
i}-\frac{2\l}{3}\Sigma_{\g \d
i}+\Psi_{ij}\Sigma_{\g \d}^j \right )+\m \Psi^k_k \label{8.13} \ee
where the independent fields now are $\Sigma^i_{\a \b}$, $A_\a ^i$, $\Psi_{ij}$
and
$\m$. Note that $\Psi_{ij}$ is an arbitrary symmetric $SO(3)$ matrix-field, its
tracelessness is imposed by the variation of $\m$. This is the Plebanski
Lagrangian,
and in order to extract real general relativity from it, one must impose
reality
conditions on $\Sigma$ that turns out to be rather awkward for Lorentzian
spacetime. I
will not go into more details concerning this action, instead I want to show
how to
reach the CDJ-Lagrangian from here.\\ \\

 The idea is the following; since the Lagrangian
now has an inhomogeneous dependence on $\Sigma$, it is possible to eliminate it
from
the action, through its equation of motion. And once this is done, the
Lagrangian
instead gets an inhomogeneous dependence on $\Psi$ making it possible to
eliminate also
this field, leaving only the curvature $F_{\a \b}^i$ and the Lagrange
multiplier $\m$.
For simplicity, I will put $\l=0$. For a treatment of the $\l\neq 0$ case,
see \cite{12b}. See also \cite{Koshti}.

\be \frac{\d S}{\d \Sigma^i_{\a \b}}=i\e^{\a \b \g \d}\left (F_{\g \d i}
+2\Psi_{ij}
\Sigma_{\g \d}^j\right )=0 \label{8.14} \ee
Assuming $\Psi_{ij}$ to be invertible, the solution is

\be \Sigma^i_{\a \b}=-\frac{1}{2}\Psi^{-1}{}^{ij}F_{\a \b j} \label{8.15} \ee
With this solution, the Lagrangian becomes

\be \cl=-\frac{i}{4}Tr(\Psi^{-1}\O)+\m Tr\Psi \label{8.16} \ee
where I have introduced a convenient matrix notation, and defined:
$\O^{ij}:=\e^{\a \b
\g \d}F_{\a \b}^iF_{\g \d}^j$. Now, using the characteristic equation for
three-by-three matrices:

\be A^3-A^2TrA+\frac{1}{2}A\left((TrA)^2-TrA^2\right )-1\;detA=0 \nn \ee
it follows that

\be Tr\Psi=\frac{1}{2}det\Psi \left ((Tr\Psi^{-1})^2-Tr\Psi^{-2}\right )
\label{8.17}
\ee
Putting this into the Lagrangian, and varying the action with respect to
$\Psi^{-1}$
one gets:

\be \frac{\d S}{\d
\Psi^{-1}_{ij}}=i\left (-\frac{1}{4}\O^{ij}+\rho(Tr\Psi^{-1}\d^{ij}-\Psi^{-1}
{}^{ij})
\right )=0 \label{8.18} \ee
where I have redefined the Lagrange multiplier: $\rho:=-i\; \m det\Psi$.
This equation is
now easily solved for $\Psi^{-1}$:

\be \Psi^{-1}{}^{ij}=-\frac{1}{4\rho}\left
(\O^{ij}-\frac{1}{2}\d^{ij}Tr\O\right )
\label{8.19} \ee
And finally, putting this solution into the Lagrangian, and redefining the
Lagrange multiplier again: $\eta:=\frac{i}{4 \rho}$, the result is the
CDJ-Lagrangian

\be \cl=\frac{\eta}{8}\left (Tr\O^2-\frac{1}{2}(Tr\O)^2\right ) \label{8.20}
\ee

\chapter{Actions in (2+1)-dimensions} \markboth{Chapter \thechapter \  \  \
Actions in (2+1)-dimensions}{\thesection \ \ \ The E-H and the H-P Lagrangian,
and
$\dots$} \label{(2+1)}

\begin{picture}(158,100)(1,20)

\thicklines
\put (109,20){\framebox(50,20){\shortstack{The Einstein-Hilbert\\ Lagrangian
${\cal L}(e^\alpha _I)$}}}
\put (2,20){\framebox(50,20){\shortstack{The Hilbert-Palatini\\ Lagrangian
${\cal
L}(e^\alpha _I,\omega ^{JK}_\beta)$}}}
\put (2,90){\framebox(50,20){\shortstack{The Ashtekar Hamiltonian\\ $\ch
(A_{aI},\p ^{bJ})$\\ $\left (A_{a I}=\e_{IJK}\o _a ^{JK}\right )$}}}
\put (109,90){\framebox(50,20){\shortstack{The ADM-Hamiltonian\\ $\ch (e^a_I,\p
_b^J)$}}}
\put(60,50){\framebox(46,20){\shortstack{The CDJ-Lagrangian\\ $\cl (\eta,
A_{\a I})$\\ $\left (A_{\a I}=\e_{IJK}\o _\a ^{JK}\right )$}}}

\thinlines
\put (52,30){\vector(1,0){57}}
\put(134,40){\vector(0,1){50}}
\put(134,80){\vector(0,-1){40}}
\put(52,100){\vector(1,0){57}}
\put(109,100){\vector(-1,0){57}}
\put(27,40){\vector(0,1){50}}
\put(52,35){\vector(4,3){20}}
\put(75,70){\vector(-4,3){27}}
\put(53.4,86.2){\vector(4,-3){20}}

\it
\put(62,25){$\frac{\d S}{\d \o}=0 \Rightarrow \o=\o (e)$}
\put(139,62){\shortstack[l]{Legendre\\ transform}}
\put(139,83){ $\p :=\frac{\d L}{\d \dot{e}}$}
\put(139,45){$\dot{e}=\{e,H\}$}
\put(5,75){\shortstack[r]{Legendre\\ transform\\ $\p :=\frac{\d L}{\d
\dot{\o}}$}}
\put(62,80){\shortstack[l]{Legendre\\ transform}}
\put(70,75){$\dot{A}=\{A,H\}$}
\put(75,40){\shortstack[l]{$\frac{\d S}{\d e}=0\Rightarrow$\\ $ e=e(\o)$}}

\put(55,108){Canonical transformation }
\put(55,103){$A_{aI}:=K_{aI}+\G _{aI}$}
\put(75,95){$K_{aI}:=A_{aI}- \G _{aI}$}
\put(35,80){$\p :=\frac{\d L}{\d \dot{A}}$}

\end{picture} \\
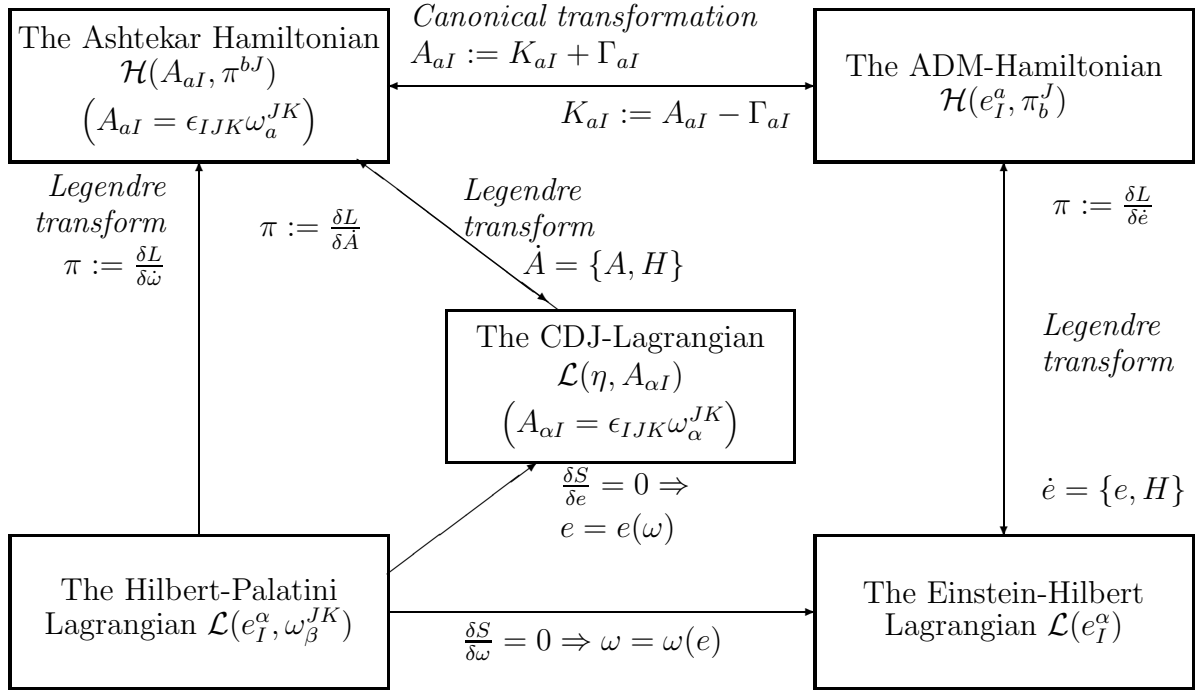
\begin{center}{\it Fig.2 Actions for gravity in (2+1)-dimensions.}\end{center}

\noindent \raisebox{-3.6mm}{\Huge I} \vspace*{-1.5mm}
n this section, I intend to briefly go through the (2+1)-dimensional
counterparts of
\hspace*{3.5mm}
the various actions dealt with in section \ref{(3+1)}. Here, I will not do all
calculations in the same detail as for (3+1)-dimensions.

The reason why (2+1)-dimensions here is treated specially, is that both the
Ashtekar
Hamiltonian as well as the CDJ-Lagrangian are known to exist only in (2+1)- and
(3+1)-dimensions. See, however, section \ref{Hd} regarding speculations about
higher dimensional formulations.
The (2+1)-dimensional version of fig.1 is given in fig.2, where we
see that the number of different actions now has decreased significantly. Two
actions
are absent, due to the non-existence of self dual two-forms in
(2+1)-dimensions, and
four of the actions from fig.1 have merged into two; the pure $SO(1,2)$
spin-connection Lagrangian equals here the CDJ-Lagrangian, and the Hamiltonian
formulation of the H-P Lagrangian equals the Ashtekar Hamiltonian.

There are at least one action missing in fig.2; the
Chern-Simons Lagrangian \cite{22}, \cite{25b}. The reason why I do not treat
the Chern-Simons Lagrangian here is that it is a
purely (2+1)-dimensional formulation, which does not exist in other dimensions,
and I
am mainly interested in formulations that exist in (3+1)-dimensions, as well.

\section{The E-H and the H-P Lagrangian, and the ADM-Hamiltonian}
\markboth{Chapter
\thechapter \ \ \
Actions in (2+1)-dimensions}{\thesection \ \  \
 The E-H and the H-P Lagrangian, and $\dots$}

All these three action formulations of gravity work perfectly all right in
arbitrary
spacetime dimensions (dimensions greater than (1+1), in (1+1)-dimensions the
E-H
and the H-P Lagrangian
are just total divergencies). All calculations that were made in sections
(\ref{EH})-(\ref{ADM}), except one, are trivially generalized to other
dimensions. The
calculation that needs a slight modification is the variation of the H-P
Lagrangian with respect to the spin-connection. In arbitrary spacetime
dimensions, the
variation yields

\be \frac{\d S_{HP}}{\d \o _\a ^{IJ}}=\cd _\b \left (e e^\a _{[I}e^\b
_{J]}\right )=0
\label{10.1} \ee
And in dimensions higher than (3+1) it is a bit tricky to show that
(\ref{10.1})
implies the zero-torsion condition (It is, however, true that it does):

\be \cd_{[\a}e_{\b ]}^I=0 \label{10.2} \ee
In (2+1)-dimensions, however, (\ref{10.2}) follows directly from (\ref{10.1}),
if one
knows the "inverse-formula":

\be ee^\a _{[I}e^\b _{J]}=\e ^{\a \b \g}\e_{IJK}e_{\g}^K \label{10.3} \ee

\section{The self dual H-P and the Plebanski Lagrangian} \markboth{Chapter
\thechapter
\ \  \
Actions in (2+1)-dimensions}{\thesection  \  \  \
 The self dual H-P and the Plebanski Lagrangian}

These two Lagrangians are unique for (3+1)-dimensions, since the construction
of them
relies heavily on the use of self dual two-forms, which only exist in four
dimensional
spacetimes. One could of course construct a Plebanski-like Lagrangian without
relation
to self duality; let the combination $e_{[\a}^Ie_{\b ]}^J$ become an
independent
field, and add a Lagrange multiplier term imposing the original definition.

\section{Hamiltonian formulation of the H-P Lagrangian, and the Ashtekar
Hamiltonian} \markboth{Chapter \thechapter \ \ \
Actions in (2+1)-dimensions}{\thesection \ \ \ Hamiltonian formulation of the
H-P
Lagrangian $\dots$} \label{HHP(2+1)}

Here is one of the formulations where (2+1)-dimensions is special. It will be
shown
here that it is exactly in (2+1)-dimensions that the Hamiltonian formulation of
the
H-P Lagrangian does not give rise to second class constraints. In fact, here in
(2+1)-dimensions, the Hamiltonian formulation equals the Ashtekar formulation.
The
reason why no second-class constraints appear here, as they did in
(3+1)-dimensions,
can be found from a counting of degrees of freedom: as we saw in section
(\ref{HHP}),
 the
second class constraints were needed since the tetrad did not have enough
independent
components to function as an unrestricted momenta for the spin-connection. In
an
arbitrary spacetime with dimension $(D+1)$, the spatial restriction of the
spin-connection has $D\frac{(D+1)D}{2}$
number of algebraically independent components, while the "tetrad" has
$(D+1)^2$
components. Then, $(D+1)$ of the "tetrad"'s components are needed as Lagrange
multipliers, imposing the diffeomorphism constraints. This means that, in order
not to
get restrictions on the momenta, we must have:

\be D\frac{(D+1)D}{2} \leq (D+1)^2 -(D+1) \ee
and the solution is $D\leq 2$, singling out (2+1)-dimensions as the unique
spacetime
where the H-P Lagrangian will not produce second class constraints. (The reason
why I
say that the additional constraints are second-class, is that a first-class
constraint
always corresponds to a local symmetry. And the H-P Lagrangian has only gauge
and
diffeomorphism symmetries.)

The standard H-P Lagrangian is

\be \cl _{HP}=\left (e e^\a _I e^\b _J R_{\a \b}{}^{IJ}(\o)+\l e\right)
\label{11.2}
\ee
where $e^\a _I$ is the triad field, $e$ the determinant of its inverse, and
$\o_\a
^{IJ}$ an independent $SO(1,2)$ connection. The Hamiltonian formulation of
(\ref{11.2}) has previously been studied in \cite{22}, \cite{10}, and
in \cite{23}. See also \cite{Geza}.
Now, I split the triad

\be e^0_I=-\frac{N_I}{N},\hspace{5mm}e^a_I=V^a_I+\frac{N^aN_I}{N};\hspace{5mm}
N>0,
\hspace{5mm}N^IN_I=-1,\hspace{5mm} V^a_IN^I=0 \label{11.3} \ee
Putting this in (\ref{11.2}) and defining: $A_\a^I:=\e^{IJK}\o_{\a JK}$, $F_{\a
\b}{}^I:=\e^{IJK}R_{\a \b JK}$, gives

\be \cl_{HP}=-\frac{e}{N}N_IV^a_J\e^{IJK}F_{0aK}+\frac{1}{2}eV^a_IV^b_J\e^{IJK}
F_{abK}
+\frac{e}{N}N^a N_IV^b_J\e^{IJK}F_{abK} + \l e \label{11.4} \ee
The momenta conjugate to $A_\a^I$ is

\be \p^{aI}:=\frac{\d L}{\d \dot{A}_{aI}}=-\frac{e}{N}N_KV^a_J\e^{KJI}
\label{11.5} \ee
Now, using the epsilon-delta identity $\e^{ab}\e_{cd}=\d^{[a}_c\d^{b]}_d$, and
the
orthogonality between $N^I$ and $V^{aI}$ it follows
that

\be \p^{aI}=-\frac{e}{N}\e^{ab}V_b^I\left(\frac{1}{2}N_J
V^c_KV^d_M\e^{JKM}\e_{cd}\right) \label{11.6} \ee
where $V_a^I$ is the inverse to $V^{aI}$: $V_{aI}V^{bI}=\d^b_a$, and
$V_{aI}N^I=0$.
Eq. (\ref{11.3}) also implies:

\be e:=(\frac{1}{6}\e_{\a \b \d}\e_{IJK}e^{\a I}e^{\b J}e^{\d
K})^{-1}=(-\frac{1}{2N}
N_J V^c_KV^d_M\e^{JKM}\e_{cd})^{-1} \label{11.7} \ee
which makes it easy to invert (\ref{11.6}):

\be V_a^I=\e_{ba}\p^{bI} \label{10.7} \ee
Putting this into the Lagrangian, and rescaling the lapse function:
$\tilde{N}:=N^2/e$, gives

\bea \cl&=&\p^{aI}\dot{A}_{aI}-\tilde{N}\ch-N^a\ch _a +A_{0I}\cg ^I
\label{11.8} \\
\ch&:=&\frac{1}{2}\p^{aI}\p^{bJ}F_{ab}{}^K\e_{IJK}-\l det(\p^{aI}\p^b_I)\approx
0 \nn \\
\ch _a&:=& \p^b_IF_{ab}{}^I\approx 0 \nn \\
\cg ^I&:=&\cd _a\p^{aI}=\partial_a\p^{aI}+\e^{IJK}A_{aJ}\p^a_K\approx 0 \nn
\eea
which is the (2+1)-dimensional Ashtekar Hamiltonian. Compare with (\ref{5.38}).
Note
that, for non-degenerate metrics, there is a unique solution to the constraints
$\ch$
and $\ch _a$: $\e ^{ab}F_{ab}^I=-\l \e ^{IJK}\p ^{a}_J\p ^{b}_K\e _{ab}$. So,
instead
of the constraints $\ch$ and $\ch _a$ given in (\ref{11.8}), one can use the
constraint: $\Psi ^I:=\e ^{ab}F_{ab}^I+\l \e ^{IJK}\p ^{a}_J\p ^{b}_K\e
_{ab}\approx
0$.

The constraint algebra for the Hamiltonian in (\ref{11.8})
	is exactly the same as for the
(3+1)-dimensional case, and there are no reality conditions here in
(2+1)-dimensions.
Following the fields through the Legendre transform above, it is easy to write
down
the metric formula for the Ashtekar variables:

\be \tilde{g}^{\a \b}=\sqrt{-g}g^{\a \b}=\left(\begin{array}{cc}
-\frac{1}{\tilde{N}}& \frac{N^a}{\tilde{N}}\\
\frac{N^a}{\tilde{N}}& \tilde{N}\p^{aI}\p^b _I -\frac{N^aN^b}{\tilde{N}}
\end{array} \right) \label{11.9} \ee
Then, it is only the relation between the Ashtekar Hamiltonian and the
ADM-Hamiltonian
left here to show. These two Hamiltonians are related by a canonical
transformation of
similar type as in the (3+1)-dimensional case. First I define the field
$K_{aI}$:

\be K_{aI}:=A_{aI}-\G_{aI} \label{11.11} \ee
where $\G_{aI}$ is the unique torsion-free spin-connection compatible with
$\p^{aI}$:

\be D_a\p^{bI}:=\partial_a\p^{bI}+\G_{ac}^b\p^{cI} -\G_{ac}^c\p^{bI}
+\e^{IJK}\G_{aJ}\p^b_K=0 \label{11.12} \ee
Despite the fact that $\p^{aI}$ does not have an inverse "in the $SO(1,2)$
indices",
(\ref{11.12}) can be uniquely solved for $\G_{aI}$ as a function of $\p^{aI}$.
This can
be understood from a counting of degrees of freedom in (\ref{11.12}): equation
(\ref{11.12})
represents 12 equations, and $\G_{ab}^c$ and $\G_{aI}$ are $6+6$ unknown,
meaning that
there is enough information in (\ref{11.12}) to solve for all components. See
Appendix
B for details. (Note, however, that the alternative requirement:
$D_{[a}\p_{b]}^I=0$, is
not enough to uniquely specify $\G_{aI}$, since it gives only 3 equations for 6
unknown.)

Now, using (\ref{11.11}) and (\ref{11.12}), the Ashtekar constraints in
(\ref{11.8})
becomes:

\bea \cg^I&=&D_a\p^{aI}+\e^{IJK}K_{aJ}\p^{a}_K=\e^{IJK}K_{aJ}\p^{a}_K
\label{11.13} \\
\ch
_a&:=&\p^{bI}R_{abI}(\G)+\p^{bI}D_{[a}K_{b]I}+\p^{bI}\e_{IJK}K_a^JK_b^K\approx
\p^{bI}D_{[a}K_{b]I} \label{11.14} \\
\ch &=& \frac{1}{2}\p^{aI}\p^{bJ}R_{ab}{}^K\e_{IJK}-\l det(\p^{aI}\p^b_I)
+\p^{aI}\p^{bJ}(D_aK_b^K)\e_{IJK} + \frac{1}{2}\p^{aI}\p^{bJ}K_{[aI}K_{b]J} \nn
\\ &\approx &
\frac{1}{2}R(\G)-\l det(\p^{aI}\p^b_I)+ \frac{1}{2}\p^{aI}\p^{bJ}K_{[aI}K_{b]J}
\label{11.15} \eea
where $R_{ab}{}^I:=\partial_{[a}\G_{b]}^I+\e^{IJK}\G_{aJ}\G_{bK}$ and
$R:=\e_{IJK}
\p^{aI}\p^{bJ}R_{ab}{}^K$, and I have neglected terms proportional to $\cg ^I$
in both
(\ref{11.14}) and (\ref{11.15}). In (\ref{11.14}) the Bianchi identity:
$\p^{bI}R_{abI}=0$ was also used. Thus, the transform (\ref{11.11}) really
gives the
wanted ADM-Hamiltonian, and what is left to prove is that the transformation
really is
a canonical one. To do that, I first define the undensitized $\p^{aI}$ and its
inverse:

\bea e^{aI}&:=&\frac{1}{\sqrt{det(\p^{aI}\p^b_I)}}\p^{aI} \nn \\
e_{aI}e^{bI}&=&\d^a_b, \hspace{10mm} \e^{IJK}e_{aI}e^b_Je^c_K=0 \label{11.16}
\eea
Then (\ref{11.12}) gives

\be D_{[a}e_{b]I}=\partial_{[a}e_{b]I}+\e_{IJK}\G_{[a}^Je_{b]}^K=0
\label{11.17} \ee
or

\be \e ^{ab}\left ( D_{a}\dot{e}_{b}^I+\e^{IJK}\dot{\G}_{aJ}e_{bK}\right )=0
\label{11.18} \ee
Then, projecting (\ref{11.18}) along the vector field
$\e^{IJK}e^a_Je^b_K\e_{ab}$,
gives

\be e^{aJ}\dot{\G}_{aJ}=2\e^{IJK}e^a_Je^b_KD_a\dot{e}_{bI} \label{11.19} \ee
which means that

\be
\p^{aI}\dot{A}_{aI}=\p^{aI}\dot{K}_{aI}+\partial_a\left(\frac{2}{\sqrt{det(e^{aI}
e^b_I)}}\e^{IJK}e^a_Je^b_K\dot{e}_{bI}\right) \label{11.110} \ee
showing that, up to the surface term, $\p^{aI}$ and $K_{aI}$ are conjugate
variables.
So, assuming compact spacetime or fast enough fall-off behavior at infinity, we
have
a canonical transformation.

\section{The CDJ-Lagrangian} \markboth{Chapter \thechapter \ \ \
Actions in (2+1)-dimensions}{\thesection \ \ \ The CDJ-Lagrangian}
\label{CDJ(2+1)}

The CDJ-Lagrangian in (2+1)-dimensions can be found either by a Legendre
transform
from the Ashtekar Hamiltonian, or from an elimination of the triad field from
the H-P
Lagrangian. The latter method is the simpler one.

\be \cl_{HP}=\e^{\a \b \g}e_{\g I}F_{\a \b }{}^I +\l e
\label{12.1} \ee
where $F_{\a \b}{}^I:=\partial_{[\a}A_{\b ]}^I+\e^{IJK}A_{\a J}A_{\b K}$, and
$A_\a^I:=\e^{IJK}\o_{\a JK}$ and $\o_{\a JK}$ is an $SO(1,2)$ connection. The
equation
of motion following from variation of $e_{\a I}$ is

\be \frac{\d S}{\d e_{\a I}}=F^{*\a I}+\l e e^{\a I}=0 \label{12.2} \ee
where $F^{*\a I}:=\e^{\a \b \g}F_{\b \g}{}^I$ is the dual of $F_{\a \b I}$. The
idea
is now to solve (\ref{12.2}) for $e^{\a I}$, and then put the solution back
into the
Lagrangian, yielding a totally metric-free formulation.

Taking the determinant of (\ref{12.2}), gives

\be \l^3e^2=-det(F^{*\a I}) \label{12.3} \ee
which is solved by

\be e=\pm\sqrt{-\frac{1}{\l^3}det(F^{*\a I})} \label{12.35} \ee
Then, (\ref{12.35}) and (\ref{12.2}) give the complete solution to
$\frac{\d S}{\d e_{\a I}}=0$:

\be e^{\a I}=\mp\frac{1}{\l \sqrt{-\frac{1}{\l^3}det(F^{*\a I})}} F^{*\a I}
\label{12.4} \ee
and putting this solution back into the Lagrangian, one gets

\be \cl=\mp 2\; sign(\l)\sqrt{-\frac{1}{\l}det(F^{*\a I})} \label{12.5} \ee
which is the wanted pure spin-connection Lagrangian for (2+1)-dimensional
gravity with
a cosmological constant. Note that this Lagrangian is totally metric-free, the
only
independent field is the spin-connection. For a treatment of the CDJ-Lagrangian
without a cosmological constant, and with a coupling to a scalar field, see
\cite{23}. See also \cite{24}.

The metric formula in this formulation, follows directly from (\ref{12.4}).

\be g^{\a \b}=e^{\a I}e^\b_I=\frac{-\l}{det(F^{*\a I})} F^{* \a I}F^{* \b}{}_I
\label{12.6} \ee
Note that the Lorentz-signature condition $detg_{\a \b}=-e^2<0$ corresponds to
\\
$\frac{1}{\l}det(F^{* \a I})<0$. See for instance (\ref{12.3}).

\chapter{Generalizations}
\markboth{Chapter \thechapter \ \ \
 Generalizations}{\thesection \ \ \ The
cosmological constants}
\label{Gen}
\raisebox{-3.6mm}{\Huge I} \vspace*{-1.5mm}
n this section, I want to present two recent generalizations of the Ashtekar
Hamiltonian \hspace*{3.5mm}
and the CDJ Lagrangian. The first kind of generalization was first
discovered in \cite{19} and in \cite{18}. See also \cite{17} and \cite{16b}. It
is
a cosmological constant type of generalization in the sense that it does not
increase
the number of degrees of freedom. At the canonical level (the Ashtekar
formulation),
the generalization is achieved by adding terms to the Hamiltonian constraint,
and at
the Lagrangian level (the CDJ Lagrangian) one has to add terms to the
Lagrangian. This
generalized theory still has an interpretation in terms of Riemannian geometry,
and it
is possible to extract the spacetime metric out of the constraint algebra.
Explicit
spherically symmetric solutions to the generalized theory have been studied in
\cite{18} and
\cite{25}, in which it is shown that the generalized theory really is
physically different from the conventional Einstein theory. Two remaining
problems
with this generalization are the reality condition and the metric-signature
requirement. No one has yet given a reality condition that works for any of the
new
cosmological constants. (The problem arises for Lorentzian spacetimes.
For Euclidean signature, where all fields are real, it is
possible to show that there exist ranges for some of the new cosmological
constants
that will ensure positive definiteness of the metric \cite{26}.)
In \cite{23}, it was also shown that this generalization
of Ashtekar variables does not have
any direct counterpart in (2+1)-dimensions. In (2+1)-dimensions the ordinary
cosmological
constant seems to be the unique generalization of the pure gravity theory, that
does
not increase the number of degrees of freedom.

The other type of generalization that I will present here is a generalization
of the
theory to other gauge groups. That is, in (3+1)-dimensions,
I consider a canonical
formulation of a gauge and diffeomorphism invariant theory that reduces to the
Ashtekar formulation if the gauge group is chosen to be $SO(3)$. See \cite{20}.
 In (2+1)-dimensions, this type of generalization is also possible, but there
the
pure gravity gauge group is $SO(1,2)$.
The difference between (3+1)- and (2+1)-dimensions is that
in (3+1)-dimensions there exist an infinite number of different theories that
has the
above mentioned behavior, while in (2+1)-dimensions there is only one theory of
this
type. This is closely related to the existence of the infinite number of new
cosmological constants in (3+1)-dimensions. This gauge group generalization
does
increase the number of degrees of freedom, and in fact has the correct number
of
degrees of freedom to be a candidate theory for Einstein gravity coupled to
Yang-Mills
theory. In (2+1)-dimensions, it was been shown in \cite{24} that this
generalization
gives the conventional Einstein-Yang-Mills theory to lowest order in Yang-Mills
fields.

In (2+1)-dimensions, there exists an extension of the pure gravity formulation,
which
has one degree of freedom per point in spacetime; topologically massive gravity
\cite{29b}. See also \cite{23} for the relation to the Ashtekar formulation. I
do not
treat this extension here, mainly due to the fact that this is a purely
(2+1)-dimensional formulation which has no counterpart in other spacetime
dimensions.

\section{The cosmological constants} \markboth{Chapter \thechapter \ \ \
Generalizations}{\thesection \ \ \
 The cosmological
constants} \label{cc}

In a canonical formulation of a diffeomorphism invariant theory, there exists a
set
of first class constraints generating the diffeomorphism symmetry. If these
generators
are split into parallel and orthogonal parts, $\tilde{\ch} _a$ and $\ch$, where
$\tilde{\ch} _a$
generates spatial diffeomorphisms on the spatial hypersurface and $\ch$
generates
diffeomorphisms off the hypersurface, these first class constraints always obey
the
following constraint algebra:

\bea \{\tilde{\ch}_a[N^a],\tilde{\ch}_b[M^b]\}&=&\tilde{\ch}_a[\pounds
_{N^b}M^a]
\label{13.1} \\
\{\tilde{\ch}_a[N^a],\ch[N]\}&=&\ch[\pounds _{N^a}N] \label{13.2} \\
\{\ch[N],\ch[M]\}&=&\tilde{\ch}_a[q^{ab}(N\partial_bM-M\partial_bN)]
\label{13.3} \eea
where $\pounds _{N^a}$ denotes the Lie-derivative along the vector field $N^a$,
and
$q^{ab}$ is the spatial metric on the hypersurface. Just by requiring
path-independence of deformations of the hypersurface, it was shown in
\cite{27} that
any any canonical formulation of a diffeomorphism invariant theory, with a
metric,
must give a representation of this algebra.
Thus, this algebra can be used
in two different ways; both as a test if a given theory is diffeomorphism
invariant,
and also as a definition of what the spatial metric is in terms of the phase
space
variables. (in fact, it is possible to extract the entire spacetime metric from
the
constraint algebra and the expression for the time-evolution of the spatial
metric. See
\cite{27} for details.)

Now, given a particular set of phase space variables, and the fact that
$\tilde{\ch}_a$ should generate spatial diffeomorphisms, it is in general easy
to
find the unique realization of $\tilde{\ch}_a$ in terms of the phase space
variables.
More specifically, if $\tilde{\ch}_a$ is the generator of spatial
diffeomorphisms,
it has
the action of the Lie-derivative on the fundamental fields:

\bea \d ^{\tilde{\ch}_a}q&:=&\{q,\tilde{\ch}_a[N^a]\}=\pounds_{N^a}q
\label{13.28} \\
\d ^{\tilde{\ch}_a}p&:=&\{p,\tilde{\ch}_a[N^a]\}=\pounds_{N^a}p \label{13.29}
\eea
where $(q,p)$ denotes the phase space variables. This normally gives a uniquely
solvable system of equations, from which it is easy to solve for
$\tilde{\ch}_a(q,p)$.
As an example, study the $SO(3)$ Yang-Mills phase space, where the fundamental
fields
are $A_{ai}$ and $E^{ai}$, an $SO(3)$ connection and its conjugate momenta.
Writing
out (\ref{13.28}) and (\ref{13.29}) in details for these phase space variables
gives
an easily solvable system of equations, with the unique solution

\be \tilde{\ch}_a=\ch _a-A_{ai}\cg^i=E^{bi}F_{abi}-A_{ai}\cd_bE^{bi}
\label{13.30} \ee
All this means that in an attempt to generalize a diffeomorphism invariant
theory, the
only freedom one has lies in the Hamiltonian constraint, $\ch$. That is, since
the
theory should be diffeomorphism invariant, it must obey the algebra
(\ref{13.1})-(\ref{13.3}), meaning that $\tilde{\ch}_a$ is the generator of
spatial
diffeomorphisms, which is unique. And furthermore, since $\tilde{\ch}_a$ is the
generator of spatial diffeomorphisms, (\ref{13.1}) and (\ref{13.2}) is
automatically
satisfied if $\tilde{\ch}_a$ and $\ch$ are diffeomorphism covariant objects.
This means that in order
to generalize a diffeomorphism invariant theory, without breaking this
invariance, the
generalization must reside in $\ch$, and the requirement it has to fulfill is
that it
should be a diffeomorphism covariant object, satisfying (\ref{13.3}).

In Ashtekar variables, there is also the $SO(3)$ symmetry, which is generated
by
$\cg^i$. In the following, I will neglect this part of the algebra and just
make sure
that the generalized $\ch$ is an $SO(3)$ scalar, in order not to break this
part of
the constraint algebra. Thus, the first basic requirements a candidate $\ch$
has to
satisfy is that it should be invariant under $SO(3)$ rotations and covariant
under
spatial diffeomorphisms. The fact that it should be invariant under $SO(3)$
transformations means that it should be constructed out of gauge covariant
objects,
like $E^{ai}$ and $B^{ai}:=\e^{abc}F_{bc}^i$, with all indices properly
contracted.
And covariance under spatial diffeomorphisms just means that all spatial
indices
should be properly contracted as well. This immediately singles out four good
candidates as basic building blocks: $A:=\e_{abc}\e_{ijk}E^{ai}E^{bj}E^{ck}$,
$B:=
\e_{abc}\e_{ijk}E^{ai}E^{bj}B^{ck}$, $C:=\e_{abc}\e_{ijk}E^{ai}B^{bj}B^{ck}$,
$D:=\e_{abc}\e_{ijk}B^{ai}B^{bj}B^{ck}$. These four scalar densities all
satisfy the
above mentioned requirements, and therefore any function of them will also do
so; an
expression such as

\be \ch=f(A,B,C,D) \ee
should therefore be a rather general Ansatz for the Hamiltonian constraint. Or,
introducing parameters that serve as
 new cosmological constants, one could consider functions like:

\be f(A,B,C,D)=\a_1A + \a_2 B + \a_3 C + \a_4 D + \a_5\frac{AB}{D} + \cdots \ee
What is left to check, to ensure that the theory is diffeomorphism invariant,
is
(\ref{13.3}). I will not give the detailed calculation here, but it is shown in
\cite{17} that it is in general true that (\ref{13.3}) is satisfied with this
type
of Ansatz.

( In (2+1)-dimensions there exist only two different basic $SO(1,2)$ vector
fields: $\Psi^I:=\e^{ab}F_{ab}^I$ and $B_I:=\e_{IJK}\p^{aJ}\p^{bK}\e_{ab}$,
meaning
that the basic building blocks for $\ch$ should be: $A:=\Psi^I\Psi_I$,
$B:=\Psi^IB_I$
and $C:=B^IB_I$. But $B + \l C$ is just the conventional Ashtekar Hamiltonian
for pure
gravity with a cosmological constant, and $A$ is proportional to $B$, when
$\ch_a=0$
is satisfied. Therefore, there exists no generalization of this type
of the pure gravity theory in
(2+1)-dimensions)

At the Lagrangian level (the CDJ Lagrangian), it is even easier to discover the
generalization. The conventional pure gravity Lagrangian is

\be \cl_{CDJ}=\frac{\eta}{8}\left(Tr\Omega^2-\frac{1}{2}(Tr\Omega)^2\right)
\label{13.60} \ee
where $\Omega^{ij}:=\e^{\a \b \g \d}F_{\a \b}^iF_{\g \d}^j$ and $\eta$ is a
scalar
density of weight minus one. Now, since $\Omega^{ij}$
is a three by three matrix, there exist three independent traces: $Tr\Omega$,
$Tr\Omega^2$ and $Tr\Omega^3$. All other scalars constructed from $\Omega^{ij}$
can be
written as functions of these three traces. This follows from the
characteristic
equation for three by three matrices. Thus, the most general $SO(3)$
invariant and diffeomorphism covariant Lagrangian density constructed out of
these
building blocks, are

\be \cl=f(\eta,Tr\Omega,Tr\Omega^2,Tr\Omega^3) \label{13.7} \ee
where one just has to make sure that the Lagrangian density is a scalar density
of
weight plus one. Or, again introducing "the cosmological constants":

\be \cl=\frac{\b_1}{\eta} +\b_2Tr\Omega +\b_3\eta Tr\Omega^2 +\b_4\eta
(Tr\Omega)^2 +
\b_5\eta^2Tr\Omega^3 + \cdots \label{13.8} \ee
These two Lagrangians are both $SO(3)$ invariant and diffeomorphism covariant,
and
will therefore in general give a Hamiltonian formulation that has the required
constraint algebra (\ref{13.1})-(\ref{13.3}). (For special values of the
cosmological
constants it can, however, happen that additional second class constraints
appear. This
is what happens if one in (\ref{13.60}) changes the factor $\frac{1}{2}$ into
$\frac{1}{3}$.)

A rather remarkable fact in this cosmological constants generalization is that
the
Urbantke formula (\ref{7.11}) still holds as an expression for the spacetime
metric. This is shown by identifying the metric from the constraint algebra,
and then
following the fields through a Legendre transform. See \cite{?}.

\section{Gauge group generalization} \markboth{Chapter \thechapter \ \ \
Generalizations}{\thesection \ \ \ Gauge group
generalization} \label{gg}

Here, I just want to briefly give the basic ideas of how to generalize the
Ashtekar
formulation to other gauge groups. For further details, see \cite{20} and
VIII.

As mentioned in section (\ref{cc}), in a canonical formulation of a
diffeomorphism
invariant theory it is always possible to find first-class constraints obeying
the
algebra (\ref{13.1})-(\ref{13.3}). And the problem with the Ashtekar
Hamiltonian for
other gauge groups, is that the crucial part of the algebra (\ref{13.3}) fails
to
close. The ordinary Hamiltonian constraint for pure gravity, in Ashtekar's
variables,
is

\be \ch=\frac{i}{4}\e_{abc}\e_{ijk}E^{ai}E^{bj}B^{ck} \label{13.90} \ee
And, in order to generalize this expression to higher dimensional gauge groups
, the
$\e_{ijk}$ must be changed into some other gauge covariant object. ($\e_{ijk}$
is only
well defined for three dimensional gauge groups.) The obvious candidate is
$f_{ijk}$,
the structure constant for the Lie-algebra. ($f_{ijk}=\e_{ijk}$ for $SO(3)$.)
This
will however not work, since the Poisson bracket $\{\ch,\ch\}$ will fail to
close. The
reason why it closes for $SO(3)$ is that there exist a structure constant
identity (the
$\e-\d$-identity) for that Lie-algebra, while there for an arbitrary gauge
group does
not exist any such identity. The idea is then to eliminate the structure
constant (or
$\e_{ijk}$) from $\ch$ without changing the content of it. And to do that, one
can use
the above mentioned structure constant identity. For instance:

\bea
\ch=\frac{detB^{ai}}{detB^{ai}}\ch&=&\frac{\frac{1}{6}\e_{def}\e_{lmn}B^{dl}B^{em}B^{f
n}}
{\sqrt{detB^{ai}B^b_i}}\frac{i}{4}\e_{abc}\e_{ijk}E^{ai}E^{bj}B^{ck}\nn \\
&=&\frac{i}{4}\frac{\e_{abc}\e_{def}(E^{ai}B^{d}_i)(E^{bj}B^e_j)(B^{ck}B^f_k)}
{\sqrt{detB^{ai}B^b_i}} \eea
is a Hamiltonian constraint that works also for higher dimensional gauge
groups, and
that reduces to the Ashtekar Hamiltonian constraint for three dimensional gauge
groups.
A perhaps even simpler way of finding the generalized Hamiltonian constraint is
to use
the Ansatz

\be E^{ai}=\Psi^{ij}B^a_j \ee
in the ordinary Ashtekar Hamiltonian constraint,

\be \ch=\frac{i}{4}\e_{abc}\e_{ijk}E^{ai}E^{bj}B^{ck}=\frac{i}{4}
detB^{ai}\left((Tr\Psi)^2-Tr\Psi^2\right) \label{13.100} \ee
and then try to construct the same constraint with the building blocks:
$E^{ai}E^b_i$,
$E^{ai}B^b_i$ and $B^{ai}B^b_i$.

\bea E^{ai}E^b_ib_{ab}&=&Tr\Psi^2 \\
E^{ai}B^b_ib_{bc}E^{cj}B_j^db_{da}&=&Tr\Psi^2 \\
E^{ai}B^b_ib_{ab}&=&Tr\Psi \eea
where I have used the fact that $\ch_a=0$ means that $\Psi^{[ij]}=0$, and
$b_{ab}$ is
the inverse to $B^{ai}B^b_i$. This means that a suitable $\ch$ is

\be \ch=-\frac{i}{4}
detB^{ai}\left(\a\;
E^{ai}E^b_ib_{ab}+(1-\a)E^{ai}B^b_ib_{bc}E^{cj}B_j^db_{da}-
(E^{ai}B^b_ib_{ab})^2\right) \label{13.20} \ee
In checking the constraint algebra however, it becomes clear that it is only
two
values of $\a$ that give a closed algebra: $\a=2$ and $\a=0$. Thus, it seems
like the
"arbitrary gauge group Hamiltonian" is more restrictive than the $SO(3)$
Hamiltonian, in
which arbitrary functions of the basic building blocks give a closed algebra.

A third way of finding the generalized theory is to start from the CDJ
Lagrangian which
is trivially generalized to other gauge groups, and then perform the Legendre
transform to the Hamiltonian formulation without using any $SO(3)$ identities.
This
procedure should always give a closed algebra, for the generators of
diffeomorphisms,
since the starting point is a manifestly diffeomorphism covariant object.

One of the
still unresolved problems with this gauge group generalization, is that the
metric,
which can be read off from the constraint algebra, looks rather awkward in
terms of
the phase space variables, and therefore the reality conditions seems to be
very hard
to find. (This is also true for the cosmological constant generalization, in
section
(\ref{cc}).)

The idea behind these type of generalization is that it may be possible to find
a
unified theory of gravity and Yang-Mills theory in this way. But, before the
reality
conditions are found, no such interpretation can be given. Another problem here
in
(3+1)-dimensions, is that there exists an infinite number of different
generalizations
that all reduce to the pure gravity Ashtekar formulation when $SO(3)$ is
chosen, and
it is not an easy task to select the correct one, if there exists one. An
optimistic
expectation regarding this, is that once the correct generalization is found,
the
reality condition problem and the "metric problem" could possibly be naturally
solved.

In (2+1)-dimensions however, the corresponding generalized theory is unique,
and there
the theory really has an interpretation as gravity coupled to Yang-Mills
theory. To
lowest order in Yang-Mills fields, the generalized theory exactly coincide with
the
conventional Einstein-Yang-Mills coupling.

\section{Higher dimensions} \label{Hd} \markboth{Chapter \thechapter \ \ \
Generalizations}{\thesection \ \ \ Higher dimensions}
Here, I want to discuss the possibility of finding higher dimensional
generalizations of
the Ashtekar Hamiltonian and the CDJ-Lagrangian. I have no real results to
present
here. When the Ashtekar Hamiltonian first was found, it seemed clear that it
was a
purely (3+1)-dimensional formulation, which relied on the use of self dual
two-forms,
which only exist in four dimensional spacetimes. Later, it was shown \cite{10}
that the
Ashtekar formulation also exists in (2+1)-dimensions. The natural question is
then: is
it also possible to find an Ashtekar Hamiltonian in dimensions higher than
(3+1)? To
try to answer that question, I will define two different meanings of "higher
dimensional Ashtekar formulation".\\

First, the perhaps obvious definition is: a canonical formulation of Einstein
gravity
on the Yang-Mills phase space. If we assume that this theory has only gauge and
diffeomorphism symmetries, we can calculate the number of degrees of freedom in
a general theory of this type, and compare with the ADM-formulation of Einstein
gravity: let
the spacetime have dimension (D+1), and the Yang-Mills gauge group have
dimension $N$.
Then, the number of degrees of freedom per point in spacetime is; for the
ADM-formulation: $\frac{D(D+1)}{2}-(D+1)=\frac{(D-2)(D+1)}{2}$, and for the
"Ashtekar
formulation": $D\times N-N-(D+1)=N(D-1)-(D+1)$, where I have subtracted the
number of
first class constraints from half the number of phase space variables. (In the
ADM-formulation, the phase space coordinate is the D-dimensional symmetric
spatial
metric, and the only local symmetries are the diffeomorphism symmetries. In the
"Ashtekar formulation", the phase space coordinate is the spatial restriction
of the
gauge connection, and the symmetries are $N$ gauge symmetries and (D+1)
diffeomorphism
symmetries.) In order to have a realization of Einstein gravity on Yang-Mills
phase
space, with the above mentioned symmetries, the number of degrees of freedom
must
coincide, meaning that

\be N=\frac{D(D+1)}{2(D-1)} \ee
Checking the known results for (2+1)- and (3+1)-dimensions, gives $N=3$ in both
cases,
which is the dimensionality for $SO(1,2)$ and $SO(3)$. Now, if it should be
possible
to find this generalization, the dimensionality $N$ of the gauge group must be
a
integer value. However, it is easy to see from the equation above that there
are no
$D>3$, that will give an integer $N$. If $N$ is an integer, then it must be
possible
to factorize out the factor $(D-1)$ from the numerator. But the numerator
consists of
the product of the two subsequent integers greater than $(D-1)$, and, although
I do
not give a proof of this statement, it is obvious that this is not possible for
$D>3$.
This means that it is impossible to find a realization of Einstein gravity on
Yang-Mills phase space, with only gauge and diffeomorphism symmetries. If one
adds
further symmetries to the theory, it may of course be possible to find this
realization. Altogether this gives the conclusion that
 the obvious higher dimensional Ashtekar formulation does not
exist.\\

Another definition of a higher dimensional Ashtekar formulation is: a gauge and
diffeomorphism invariant canonical formulation on the Yang-Mills phase space.
Thus, one
just relaxes the requirement that the theory must equal the Einstein theory of
gravity, and instead allows the theory to have an arbitrary number of degrees
of
freedom. With this definition, I believe it is possible to find a higher
dimensional
Ashtekar formulation. (In fact, the conventional (3+1)-dimensional Ashtekar
Hamiltonian with gauge group $SO(3)$, is trivially generalized to higher
dimensions.
However, in dimensions higher than (3+1), the spatial metric will then always
be
degenerate.) In order to find the Ashtekar Hamiltonian one must again
concentrate on
finding a suitable Hamiltonian constraint. Gauss' law and the vector constraint
are
trivially generalized to arbitrary spacetime dimensions and gauge groups. And
again,
the requirements the Hamiltonian constraint must satisfy are that it should be
gauge
and diffeomorphism covariant, and obey (\ref{13.3}). I will not try to guess or
derive
any Hamiltonian constraints here, instead I will describe another way of
finding this
formulation. If we start at the CDJ-level, which is easier generalized to
higher dimensions, the Legendre transform will take us to the wanted
Hamiltonian. To
construct the CDJ-Lagrangian in arbitrary spacetime dimensions, one may use the
building blocks: the scalar density field $\eta$, two epsilon tensor densities,
$\e^{\a \b \g \dots}$ and (D+1) field strengths, $F_{\a \b}^I$. Then, one needs
some
gauge covariant objects like the killing-metric and/or the structure constants
from
the Lie-algebra, to properly contract the gauge-indices. This, in general
non-unique, Lagrangian is both gauge and diffeomorphism covariant, and the
Hamiltonian
formulation will therefore satisfy the required algebra
(\ref{13.1})-(\ref{13.3}). (It
could of course happen that additional second class constraints appear, as
well.)
However, the major problem with this type of construction is that the Legendre
transform generally is rather problematic to perform. (Nobody has yet showed
how this
should be done, in dimensions higher than (3+1).)

So, to summarize the speculations; it seems that the obvious generalization of
Ashtekar's variables to higher dimensions, is {\em not} possible to find,
without the
introduction of additional symmetries. However, if one just seeks a gauge and
diffeomorphism invariant formulation on Yang-Mills phase space, it ought to be
possible to find this formulation from a Legendre transform from the higher
dimensional CDJ-Lagrangian.

\chapter{Outlook} \markboth{Chapter \thechapter \ \ \ Outlook}{}
\raisebox{-3.6mm}{\Huge W} \vspace*{-1.5mm}
hen it became clear that conventional Einstein gravity is perturbatively
non-renormal- \hspace*{8.5mm} izable,
and that the attempts of improving this behavior, such as higher
derivative theories and supergravity, also had failed, many physicists took
this as a
sign that gravity and/or quantum mechanics need a drastic modification before
it will
be possible to unite them. The alternative conclusion from this failure is that
it is the methods that are wrong, and, in
fact, knowing the failure of perturbative quantization of gravity, it is easy
to find
reasons why a perturbative approach to quantum gravity is bound to fail; see
{\it e.g}
\cite{3b}, \cite{8}, \cite{31} and \cite{32} for enlightening discussions of
perturbative verses
non-perturbative treatments of quantum gravity.

Without the adequate skill of handling non-perturbative path integrals, we are
left
with canonical quantization \`{a} la Dirac, in order to be able to handle the
quantization of gravity, non-perturbatively. Earlier attempts of quantizing
gravity
canonically, have all been based on the ADM-Hamiltonian, and these attempts
have failed
due to both technical as well as conceptual problems. The technical problems
are
mostly concentrated in the complicated Hamiltonian constraint; the
Wheeler-DeWhitt
equation. Since the Hamiltonian constraint, in the ADM-formulation, has a
complicated
non-polynomial dependence on the basic phase space coordinate, the
spatial metric, no one has yet been able to find an explicit, and well-defined,
 quantum solution to this
constraint. The conceptual problems are the problems that any attempt of
quantizing
gravity eventually will have to face; {\it e.g} the problem of time, and the
problem
of finding/defining physical observables in a theory of quantum gravity
\cite{8}, \cite{31} \cite{33}, \cite{34}.

When Ashtekar \cite{2} reformulated the Hamiltonian for gravity to be a
formulation on
$SO(3)$ Yang-Mills phase space, with rather simple polynomial and homogeneous
constraints, the above mentioned technical problems were reduced significantly.
Instead, the Ashtekar Hamiltonian has another new feature that was absent in
the
ADM-Hamiltonian; the theory is complex, and one needs reality conditions in
order to
be able to extract real general relativity. This could be a serious drawback
for the
Ashtekar formulation, or as an optimist would say: the reality condition may be
something positive in that they will help us to select the correct inner
product for
the theory.

Anyway, since the discovery of the Ashtekar Hamiltonian, the quantization
program,
using Dirac quantization of the Ashtekar Hamiltonian, has grown to be an active
field
of research. There already exist two excellent books covering this subject
\cite{8},
\cite{31}, a number of different reviews \cite{35}, \cite{36}, \cite{37},
\cite{38},
and more than 300 publications \cite{39}, all related to Ashtekar's variables.

The first attempts, in this program, used the so called connection
representation
\cite{3a}, where the wave functionals are holomorphic functionals of the self
dual
Ashtekar connection, and the trace of the holonomy of this connection around a
closed
loop can be shown to satisfy both Gauss' law and the Hamiltonian constraint.
The
vector constraint is not solved, however. This soon led to the
loop-representation, where the wave functionals are complex functionals of
loops
on the spatial hypersurface \cite{3b}, \cite{36b}, \cite{40}, \cite{41}. In
this
representation, the algebra that is quantized is a non-canonical graded
$T$-algebra,
where $T$ stands for traces of holonomies with momentum fields inserted along
the
loop. This algebra is automatically $SU(2)$ ($SO(3)$) invariant, which means
that
Gauss' law is already taken care of at the classical level (reduced phase space
quantization). The vector constraint, which classically generates spatial
diffeomorphisms, is solved \`{a} la Dirac by only considering wave functionals
of
knot and link classes of loops. (One really considers generalized knot and link
classes, for details see {\em e.g} \cite{38b}.) It was also soon clear that the
Hamiltonian constraint is solved by using only non-intersecting, smooth loops
\cite{3b}. Later, several other solutions to all the quantum constraints have
been
found \cite{40}, \cite{41}. However, there are still two extremely important
ingredients missing; the inner product and observables. Without these, one
cannot calculate any physical quantities, and the solutions do not really give
any
information about the theory. Another important ingredience is the {\em
general}
solution to all the constraints. Perhaps that is too much to ask for, but what
is
really needed is an understanding of the importance of the known solutions. For
all we
know, the solutions that have been found so far could all belong to a
"degenerate set
of measure zero". (That is, they could be totally unimportant.)

The quantization program also includes
work on linearized gravity \cite{42}, \cite{43}, (2+1)-dimensional gravity
\cite{2+1},
 Maxwell fields \cite{44}, \cite{44b},
(1+1)-dimensional QED \cite{45}, etc.\\ \\
Finally, I want to give my opinion and expectations of how the CDJ-formulation
as
well as the generalizations of Einstein gravity can contribute
to a better understanding of classical and quantum gravity. So far,
nothing really important has come out of the discovery of the CDJ-Lagrangian.
(When
the CDJ-Lagrangian was found \cite{12}, the general solution to the
classical diffeomorphism constraints, modulo Gauss' law, were also found, but
this
could have been found without the knowledge of
the CDJ-Lagrangian.) There have been attempts to make progress with path
integral
quantization, and discretized approximations
 using the CDJ-Lagrangian \cite{46}, \cite{47}, but so far there are no
real results coming out of this. Related to the CDJ-Lagrangian there have also
been
some new discoveries concerning gravitational instantons \cite{48}, \cite{49},
\cite{Torre}. In my
opinion, the most interesting outcome of the CDJ-Lagrangian is, so far, the
gauge
group generalization. This generalization could also have been found without
the
knowledge of the CDJ-formulation, but it is really techniques and ideas coming
from
this pure connection formulation that have enabled the finding of this
generalization.

Why is the generalizations, presented in section 4, so interesting?
So far, none of these
generalizations has a satisfactory treatment of the reality conditions, or the
metric
signature condition. However, if these, and possibly other, problems can be
solved,
the new cosmological constants are on the same footing as the conventional
Einstein
cosmological constant, and we will need a theoretical explanation of why they
are all
zero, or keep them in the theory. Perhaps, the introduction of all these new
cosmological constants will eventually help us understand why it is the pure
gravity
Einstein
equation without any cosmological constants that seems to be the correct
equation for
describing "classical nature".
A positive expectation regarding the gauge group generalization of Ashtekar's
variables, is that this generalization may be a suitable theory for the
loop-representation quantization of coupled gravity-Yang-Mills theory. This is
otherwise a problem in the Ashtekar formulation; the coupling of gravity to
Yang-Mills
theory has a Hamiltonian constraint that is
seemingly not suited for the loop-representation quantization.

\appendix

\chapter{Conventions and Notation} \markboth{Appendix A}{Conventions and
Notation} \label{A}

\underline{Indices}: $\a , \; \b , \; \g , \dots$ denote spacetime indices,
$I$,
$J$, $K$, $\dots$ denote $SO(1,3)$ indices (or $SO(1,2)$ in (2+1)-dimensions),
$a$,
$b$, $c$, $\dots$ denote spatial indices.\\ \\
\underline{Symmetrization and Antisymmetrization}: Symmetrization and
antisymmetrization of indices are denoted by brackets, according to

\bea A^{(ab)}&:=&A^{ab}+A^{ba} \\
A^{[ab]}&:=&A^{ab}-A^{ba} \\
A^{(abc)}&:=&A^{(ab)c}+A^{(ca)b}+A^{(bc)a} \\
A^{[abc]}&:=&A^{[ab]c}+A^{[ca]b}+A^{[bc]a} \\
A^{(a|b|c)}&:=&A^{abc}+A^{cba} \eea
\\ \\
\underline{The Minkowski metric}: The Minkowski metric is chosen to be
$\eta^{IJ}:=Diag(-1,1,1,1)$ in (3+1)-dimensions, and $\eta^{IJ}:=Diag(-1,1,1)$
in
(2+1)-dimensions.
\\ \\
\underline{(3+1)-dimensional $\e$ -symbol}: $\e^{\a \b \g \d}$ and $\e_{\a \b
\g \d}$
are totally antisymmetric, and $\e^{0123}=\e_{0123}=1$ in every coordinate
system,
implying that $\e^{\a \b \g \d}$ is a tensor density of weight plus one, and
$\e_{\a \b \g \d}$ has weight minus one. For any non-degenerate tensor $K_{\a
\b}$,
the following is true

\be \e_{\a \b \g \d}=\frac{1}{K}K_{\a \e}K_{\b \s}K_{\g \r}K_{\d \k}\e^{\e \s
\r \k}
\ee
where $K:=detK_{\a \b}=\frac{1}{24}\e^{\a \b \g \d}\e^{\e \s \r \k} K_{\a
\e}K_{\b \s}
K_{\g \r}K_{\d \k}$.
\\ \\
\underline{Spatial restriction of $\e^{\a \b \g \d}$}: $\e^{abc}:=\e^{0abc}$,
$\e_{abc}
:=\e_{0abc}$, meaning that $\e^{abc}$ satisfies the following identities

\bea \e^{abc}\e_{def}&=&\d^{[a}_d\d^b_e\d^{c]}_f \\
\e^{abc}\e_{ade}&=&\d^{[b}_d\d^{c]}_e \\
\e^{abc}\e_{abe}&=&2\d^c_e \eea
\\ \\
\underline{$SO(1,3)$ $\e$ -symbol}: $\e^{IJKL}$ is totally antisymmetric, and
$\e^{0123}=1$. \\ $\e_{IJKL}:=\eta_{IM}\; \eta_{JN}\; \eta_{KP}\;
\eta_{LQ}\; \e^{MNPQ}$, meaning
that $\e_{0123}=-1$, and the following identities are valid:

\bea \e^{IJKL}\e_{MNPQ}&=&-\d^{[I}_M\d^J_N\d^K_P\d^{L]}_Q \\
\e^{IJKL}\e_{INPQ}&=&-\d^{[J}_N\d^K_P\d^{L]}_Q \\
\e^{IJKL}\e_{IJPQ}&=&-2 \d^{[K}_P\d^{L]}_Q \\
\e^{IJKL}\e_{IJKQ}&=&-6\d^L_Q \eea
\\ \\
\underline{(2+1)-dimensional $\e$ -symbol}: $\e^{\a \b \g}$ and $\e_{\a \b \g}$
are totally antisymmetric, and $\e^{012}=\e_{012}=1$ in every coordinate
system,
implying that $\e^{\a \b \g}$ is a tensor density of weight plus one, and
$\e_{\a \b \g}$ has weight minus one. For any non-degenerate tensor $K_{\a
\b}$,
the following is true

\be \e_{\a \b \g}=\frac{1}{K}K_{\a \e}K_{\b \s}K_{\g \r}\e^{\e \s \r}
\ee
where $K:=detK_{\a \b}=\frac{1}{6}\e^{\a \b \g}\e^{\e \s \r} K_{\a \e}K_{\b \s}
K_{\g \r}$.\\ \\

\noindent \underline{Spatial restriction of $\e^{\a \b \g}$}:
$\e^{ab}:=\e^{0ab}$,
$\e_{ab}
:=\e_{0ab}$, meaning that $\e^{ab}$ satisfies the following identities

\bea \e^{ab}\e_{de}&=&\d^{[a}_d\d^{b]}_e\\
\e^{ab}\e_{ae}&=&\d^b_e \eea
\\ \\
\underline{$SO(1,2)$ $\e$ -symbol}: $\e^{IJK}$ is totally antisymmetric, and
$\e^{012}=1$. \\ $\e_{IJK}:=\eta_{IM}\; \eta_{JN}\; \eta_{KP}\; \e^{MNP}$,
meaning
that $\e_{012}=-1$, and the following identities are valid:

\bea \e^{IJK}\e_{MNP}&=&-\d^{[I}_M\d^J_N\d^{K]}_P\\
\e^{IJK}\e_{INP}&=&-\d^{[J}_N\d^{K]}_P\\
\e^{IJK}\e_{IJP}&=&-2\d^K_P \eea

\chapter{Definitions of Connections and Curvature} \markboth{Appendix
B}{Definitions
of Connections and Curvature} \label{B}

\underline{Definition of metric compatible affine connection and
spin-connection.} The
affine connection $\G^{\g} _{\a \b}$ and the spin-connection $\o^{IJ}_\a$ are
here
defined to annihilate the tetrad $e_{\a I}$:

\be \cd_\a e_{\b I}:=\partial_\a e_{\b I}-\G^{\g} _{\a \b}e_{\g I} + \o _{\a
I}{}^J
e_{\b
J}=0 \label{B.1} \ee
(Zero-torsion is assumed, and therefore $\G^{\g} _{[\a \b]}=0$.) This means
that

\be \cd_\a (e_{\b I}e^\b _J)=\cd_\a \eta_{IJ}=\partial_\a \eta_{IJ}+\o_{\a
I}{}^K\eta_{KJ} +\o_{\a J}{}^K\eta_{IK}=\o_{\a (IJ)}=0 \label{B.2} \ee
or, $\o_\a{}^{IJ}$ is antisymmetric, and is therefore $so(1,3)$ Lie-algebra
valued.
Now, to solve for $\G^{\g} _{\a \b}$ and $\o^{IJ}_\a$ from (\ref{B.1}), it is
convenient
to first solve for $\G^{\g} _{\a \b}$ from the equation:

\be \cd_\a g_{\b \g}=\cd_\a(e_{\b I}e_{\g} ^I)=\partial_\a g_{\b \g}-\G^\e_{\a
\b}g_{\e
\g} -\G^\e_{\a \g}g_{\b \e}=0 \label{B.3} \ee
To solve for $\G^{\g} _{\a \b}$, one does a cyclic permutation of the three
indices
$\a$, $\b$ and $\g$ two times. The two resulting equations are then added
together,
and finally equation (\ref{B.3}) is subtracted from this sum. Due to the
symmetry of
$\G^{\g} _{\a \b}$, the result is

\be \partial _\a g_{\b \g} + \partial _{\g} g_{\a \b} -\partial _\b g_{\g \a} -
2 \G^\e
_{\g \a}g_{\e \b}=0 \label{B.4} \ee
which is easily solved.

\be \G^\e _{\a \b}=\frac{1}{2}g^{\e \g}(\partial _\a g_{\e \b} + \partial _\b
g_{\e
\a} -\partial_\e g_{\a \b}) \label{B.5} \ee
Then, putting (\ref{B.5}) into (\ref{B.1}), the solution for $\o_{\a IK}$
becomes:

\be \o_{\a}^{IK}=-e^{\b K}\partial_\a e_{\b}^I + e^{\b K}\G^{\g} _{\a
\b}e_{\g}^{
I}=\frac{1}{2}e^{\e [I}\left(\partial_{[\a}e^{K]}_{\e ]} + e^{\b K]}e_\a
^L\partial_\b
e_{\e L}\right) \label{B.6} \ee
Thus, (\ref{B.5}) and (\ref{B.6}) are the unique solution to (\ref{B.1}), and
an
indication to why it is possible to uniquely solve for these connections, can
be found
by counting degrees of freedom: In (D+1)-dimensions, equation (\ref{B.1}) has
$(D+1)^3$
components, while $\G^{\g} _{\a \b}$ and $\o^{IJ}_\a$ has
$\frac{(D+1)(D+2)}{2}(D+1)$
and $\frac{(D+1)D}{2}(D+1)$ algebraically independent components, showing that
the
number of unknowns are the same as the number of equations. So, as long as the
system
of equations (\ref{B.1}) is non-degenerate, there will always exist a unique
solution.
(The non-degeneracy of (\ref{B.1}) is closely connected to the non-degeneracy
of the
tetrad field.)\\ \\
\underline{Another derivation of the spin-connection.} Instead of (\ref{B.1}),
the
spin-connection can be defined as follows:

\be \cd _{[\a}e_{\b ]I}:=\partial_{[\a}e_{\b ]I}+\o_{[\a I}{}^Je_{\b ]J} =0
\label{B.7}
\ee
Note that (\ref{B.1}) implies (\ref{B.7}) but the converse is not true.
($\G^{\g} _{\a
\b}$ is not even defined in (\ref{B.7}).) Do (\ref{B.1}) and (\ref{B.7}) then
have the
same solution for $\o_\a^{IJ}$? Yes, they have, and that can be understood by
dimensional counting again: In (D+1)-dimensions, (\ref{B.7}) represents
$\frac{(D+1)D}{2}(D+1)$ equations, and the spin-connection has the same number
of
algebraically independent components, meaning that (\ref{B.7}) has a unique
solution.
Then, since both (\ref{B.1}) and (\ref{B.7}) have unique solutions, and
(\ref{B.1})
implies (\ref{B.7}), the solution must be the same.

To directly solve (\ref{B.7}) for $\o_\a ^{IJ}$, one can use the same trick
that was
used to solve for $\G^{\g} _{\a \b}$. First, convert all free indices in
(\ref{B.7})
into
flat $SO(1,3)$ indices:

\be e^\a _Je^\b _K\left(\partial_{[\a}e_{\b ]I}+\o_{[\a I}{}^Le_{\b
]L}\right)=0
\label{B.8} \ee
Then, do a cyclic permutation of the free indices $I$, $J$ and $K$ two times,
and add
and subtract the three resulting equations:

\be \O_{JKI}+\O_{IJK}-\O_{KIJ}+2 e^\a _J\o_{\a IK}=0 \ee
where I have defined $\O_{JKI}:=e^\a _Je^\b _K \partial_{[\a}e_{\b ]I}$. The
solution
for the spin-connection is

\be \o_{\a KI}=\frac{1}{2}e_\a ^J\left(\O_{JKI}+\O_{IJK}-\O_{KIJ}\right)
\label{B.9}
\ee
Comparing (\ref{B.9}) to (\ref{B.6}) gives exact agreement. \\ \\
\underline{Definition of the curvature field.} The Riemann tensor is defined as
follows:

\bea \cd _{[\a}\cd _{\b ]}\l _\e&=&R_{\a \b \e}{}^\m \l _\m \label{B.10} \\
\cd _{[\a}\cd _{\b ]}\l _I&=&R_{\a \b I}{}^J\l _J \label{B.11} \eea
By using these definitions for a vector field of the form $\l _\e=e_\e ^I\l
_I$, one
can show that

\be e^\e _I\cd _{[\a}\cd _{\b ]}\l _\e=e^\e _I R_{\a \b \e}{}^\m e_\m ^J\l _J =
R_{\a \b I}{}^J\l _J \label{B.12} \ee
And, since this is valid for all vectors $\l _I$, the following relation must
hold:

\be R_{\a \b I}{}^J=R_{\a \b \e}{}^\m e^\e _I e_\m ^J \label{B.13} \ee
Now, using the definitions (\ref{B.10}) and (\ref{B.11}) together with the
definition
of the covariant derivative (\ref{B.1}), it is straightforward to derive the
explicit
form of the Riemann tensor:

\bea R_{\a \b}{}^\e{}_\m&=&\partial_{[\a}\G^\e _{\b ]\m} + \G^\e _{[\a \r}\G^\r
_{\b
]\m} \label{B.14} \\
R_{\a \b}^{IJ}&=&\partial_{[\a}\o_{\b ]}{}^{IJ} + \o _{[\a}{}^{IK}\o _{\b
]}{}_K{}^J
\label{B.15} \eea \\ \\

\noindent
 \underline{Einstein's equation.} Einstein's equation for pure gravity with a
cosmological constant, in the metric formulation, is

\be R_{\m \n}-\frac{1}{2}g_{\m \n}R -\l g_{\m \n}=0 \label{B.16} \ee
Using the relation (\ref{B.13}), the definition of the Ricci tensor $R_{\m
\n}:=R_{\m \a \n}{}^\a$ and the definition of the curvature scalar $R:=g^{\m
\n}R_{\m
\n}$, Einstein's equation can be rewritten in terms of the tetrad.

\be R_{\n \a I}{}^Je_\m ^I e^\a _J -\frac{1}{2}e_\m ^K e_{\n K}\left( e^{\g
I}e^\a _J
R_{\g \a I}{}^J + 2 \l\right) =0 \label{B.165} \ee
Or, multiplying with $e^{\m I}$ to get the form that follows from the
Einstein-Hilbert
Lagrangian:

\be R_{\n \a}{}^{IJ}e^\a _J -\frac{1}{2}e_\n ^I\left( e^{\g K}e^\a _J R_{\g \a
K}{}^J
+ 2\l\right) =0 \label{B.17} \ee \\ \\

\noindent
\underline{"Hybrid connections."} Here, I want to show that it is possible to
define a
unique spin-connection compatible with a sort of "hybrid" vielbein variable
$e_a ^I$,
where the dimensionality of the spatial index is one unit lower than the
dimensionality of the internal index.

Consider a $(D+1)$-dimensional spacetime, and a "vielbein" field $e_{\a I}$
where $\a$
take values $0, 1, \dots, D$, and the $I$-index is a flat Lorentz index, taking
values
$0, 1, \dots D$. In a Hamiltonian formulation where one has to partly break the
manifest spacetime covariance, the "vielbein" is split into two parts, $e_{0I}$
and
$e_{aI}$, where $a$ is a spatial index taking values $1, 2, \dots, D$. Then, I
define
a spin-connection $\o_a{}^{IJ}$ compatible with this "hybrid" field $e_{aI}$:

\be D_a e_{bI}:=\partial_ae_{bI}-\G^c_{ab}e_{cI} + \o_{aI}{}^Je_{bJ}=o
\label{B.18}
\ee
At first sight, it seems to be impossible to find a unique solution for
$\o_{aI}{}^J$,
since there exists no inverse to $e_{aI}$, such that $e^{aJ}e_{aI}=\d^J_I$. It
is
however possible to uniquely solve for both $\G^c_{ab}$ and $\o_{aI}{}^J$, and
that
can again be understood by counting degrees of freedom: Equation (\ref{B.18})
represents
$D^2(D+1)$ equations, and $\G^c_{ab}$ and $\o_{aI}{}^J$ have
$\frac{D(D+1)}{2}D$ and
$\frac{D(D+1)}{2}D$ number of algebraically independent components, giving
exact
agreement between the number of unknown and the number of equations. Another
way of
understanding this is to note that since $\o_a ^{IJ}$ is antisymmetric, one
does not
loose any information in $\o_a ^{IJ}$ by projecting it on $e_{aI}$ as long as
$e_{aI}$
represents $D$ linearly independent $SO(1,D)$ vectors.

To explicitly solve (\ref{B.18}) for $\o_a^{IJ}$, it is convenient to introduce
the
unit, time-like vector field, $N^I$, orthogonal to $e_{aI}$:

\be N^Ie_{aI}=0,\hspace{5mm}N^IN_I=-1,\hspace{5mm}\Rightarrow
N^I=\pm\frac{\e^{IJK\cdots P}e_{aI}e_{bJ}e_{cK}\cdots e_{gP}\e^{abc\cdots g}}
{D!\sqrt{
det(e_{aI}e_b^I)}} \label{B.19} \ee
Then, I define the projection operator:

\be \tilde{\eta}^{IJ}=e^{aI}e_a^J=\eta^{IJ}+N^IN^J \label{B.20} \ee
where $e^{aI}:=q^{ab}e_b^I$ and $q^{ab}$ is the inverse to
$q_{ab}:=e_{aI}e_b^I$. I
use this projection operator to project out time-like (primed) and space-like
(tilded)
indices.

\be \l^I=\l^{I'}+\l^{\tilde{I}}=-N^IN_J\l^J +\tilde{\eta}^{IJ}\l_J
=\eta^{IJ}\l_J
\label{B.201} \ee
Using this, $\o_a^{IJ}$ can be written as

\be \o_a{}^{IJ}=\o_a^{[I'\tilde{J}]}+\o_a{}^{\tilde{I}\tilde{J}} \label{B.21}
\ee
Now, it is straightforward to solve (\ref{B.18}) for $\G^c_{ab}$,
$\o_a^{[I'\tilde{J}]}$ and $\o_a{}^{\tilde{I}\tilde{J}}$. First, $\G^c_{ab}$
can be
solved for as usual:

\be \G^c_{ab}=\frac{1}{2}q^{ce}\left(\partial_a q_{eb}+ \partial_b q_{ea}
-\partial_e
q_{ab}\right) \label{B.22} \ee
and then the spin-connection is easily given:

\bea \o_{aI'\tilde{J}}&=&e^b_JN_IN^K\partial_ae_{bK} \label{B.23} \\
\o_{a\tilde{I}\tilde{J}}&=&-\tilde{\eta}_I{}^Ke^b_J\partial_ae_{bK}+e^b_J\G^c_{ab}e_
{cI} \label{B.24} \eea
Thus, (\ref{B.21}), (\ref{B.22}), (\ref{B.23}) and (\ref{B.24}) give the unique
solution to (\ref{B.18}).

Note however that the alternative definition

\be D_{[a}e_{b]I}=\partial_{[a}e_{b]I} + \o_{[a I}{}^Je_{b]J}=0 \label{B.25}
\ee
does not have a unique solution for $\o_{aI}{}^J$, which can be seen by
dimensional
counting: The number of equations are $\frac{D(D-1)}{2}(D+1)$ while the number
of
unknowns are $\frac{D(D+1)}{2}D$, showing that there are to few equations to
uniquely
specify the spin-connection. \\ \\
\underline{Variations of $\o_\a^{IJ}$ and $R_{\a \b}{}^{IJ}$.} From
(\ref{B.15}) it
follows that

\be \d R_{\a \b}{}^{IJ}=\partial_{[\a}\d \o_{\b ]}{}^{IJ} + \d \o _{[\a}{}^{IK}
\o
_{\b]K}{}^J + \o _{[\a}{}^{IK}\d \o _{\b ]K}{}^J=\cd _{[\a}\d \o_{\b ]}{}^{IJ}
\label{B.26} \ee
Note that $\d \o _\a {}^{IJ}$ is a Lorentz covariant object since it is the
difference
between two Lorentz connections. To find the variation of $\o _a {}^{IJ}$ with
respect
to variations of $e_{\a I}$, one can use the defining equation:

\be \cd _{[\a}e_{\b ]I}:=\partial_{[\a}e_{\b ]I}+\o_{[\a I}{}^Je_{\b ]J} =0
\ee
Varying this equation, gives

\be \cd _{[\a}\d e_{\b ]I} + \d \o_{[\a I}{}^Je_{\b ]J} =0 \label{B.27} \ee
But, (\ref{B.27}) is the same equation as (\ref{B.7}), with $\o _\a
{}^{IJ}\rightarrow
\d \o _\a {}^{IJ}$ and $\partial_{[\a}e_{\b ]I}\rightarrow \cd _{[\a}\d e_{\b
]I}$,
meaning that the solution to (\ref{B.27}) is

\be \d \o _\a{}^{IJ}=\frac{1}{2}e^{\e [I}\left( \cd _{[\a}\d \e_{\e ]}^{J]} +
e^{\b
J]} e_\a ^K \cd _\b \d e_{\e K}\right) \label{B.28} \ee
Together, (\ref{B.26}) and (\ref{B.28}) then give the variation of $R_{\a \b
IJ}$ with
respect to $e_{\a I}$.

\end{document}